\documentclass{aa}
\usepackage{epsfig}
\usepackage{natbib}
\bibpunct{(}{)}{;}{a}{}{,}

\newbox\grsign \setbox\grsign=\hbox{$>$} \newdimen\grdimen \grdimen=\ht\grsign
\newbox\simlessbox \newbox\simgreatbox \newbox\simpropbox \newbox\wtildebox 
\setbox\simgreatbox=\hbox{\raise.5ex\hbox{$>$}\llap
     {\lower.5ex\hbox{$\sim$}}}\ht1=\grdimen\dp1=0pt
\setbox\simlessbox=\hbox{\raise.5ex\hbox{$<$}\llap
     {\lower.5ex\hbox{$\sim$}}}\ht2=\grdimen\dp2=0pt
\def\simgreat{\mathrel{\copy\simgreatbox}}
\def\simless{\mathrel{\copy\simlessbox}}

\newcommand{\be}{\mbox{\begin{equation}}}
\newcommand{\ee}{\mbox{\end{equation}}}

\newcommand{\mmin}{\mbox{$m_{\rm min}(t)$}}
\newcommand{\mmini}{\mbox{$m_{\rm min,i}$}}
\newcommand{\mmax}{\mbox{$m_{\rm max}(t)$}}
\newcommand{\mmaxi}{\mbox{$m_{\rm max,i}$}}

\newcommand{\tdis}{\mbox{$t_{\rm dis}^{\rm total}$}}

\newcommand{\Cref}{\mbox{$m_{\rm ref}$}}

\newcommand{\qev}{\mbox{$q_{\rm ev}^{\rm lum}$}}
\newcommand{\qevt}{\mbox{$q_{\rm ev}^{\rm lum}(t)$}}
\newcommand{\qevtot}{\mbox{$q_{\rm ev}^{\rm tot}$}}
\newcommand{\qevtott}{\mbox{$q_{\rm ev}^{\rm tot}(t)$}}

\newcommand{\muevt}{\mbox{$\mu_{\rm ev}^{\rm lum}(t)$}}

\newcommand{\muevsrt}{\mbox{$\mu_{\rm ev}^{\rm sr}(t)$}}
\newcommand{\muevtot}{\mbox{$\mu_{\rm ev}^{\rm tot}$}}
\newcommand{\muevtott}{\mbox{$\mu_{\rm ev}^{\rm tot}(t)$}}

\newcommand{\msun}{\mbox{M$_\odot$}}

\renewcommand{\d}{{\rm d}} 
\newcommand{\mc}{\mbox{$M_{\rm cl}$}}
\newcommand{\mci}{\mbox{$M_{\rm cl,i}$}}

\newcommand{\mcli}{\mbox{$M_{\rm cl,i}$}}
\newcommand{\mcl}{\mbox{$M_{\rm cl}^{\rm lum}$}}
\newcommand{\mclt}{\mbox{$M_{\rm cl}^{\rm lum}$}(t)}
\newcommand{\mctot}{\mbox{$M_{\rm cl}^{\rm tot}$}}
\newcommand{\mctott}{\mbox{$M_{\rm cl}^{\rm tot}$}(t)}
\newcommand{\msr}{\mbox{$M_{\rm cl}^{\rm sr}$}}
\newcommand{\msrt}{\mbox{$M_{\rm cl}^{\rm sr}$}(t)}
\newcommand{\mssr}{\mbox{$m_{\rm sr}$}}
\newcommand{\mssrt}{\mbox{$m_{\rm sr}(t)$}}
\newcommand{\ms}{\mbox{$m_{\rm s}$}}

\begin{document}

\title{The photometric evolution of star clusters and the preferential loss of low-mass bodies - with an application to globular clusters\thanks{The models presented in this paper are publicly available in electronic form at the CDS via anonymous ftp to \texttt{http://cdsweb.u-strasbg.fr/} (130.79.128.5) or via \texttt{http://cdsweb.u-strasbg.fr/cgi-bin/qcat?J/A+A/} and also at \texttt{http://www.astro.uu.nl/\~{}kruijs}.}
}

\author{J.~M.~Diederik~Kruijssen \inst{1,2} and
        Henny~J.G.L.M.~Lamers \inst{1}}

\institute { 
                 {\inst{1}Astronomical Institute, Utrecht University, 
                 Princetonplein 5, NL-3584CC Utrecht, The Netherlands\\
                 (e-mail: {\tt kruijssen@astro.uu.nl; lamers@astro.uu.nl})}\\
                 {\inst{2}Sterrewacht Leiden, Leiden University,
                 PO Box 9513, NL-2300RA Leiden, The Netherlands}
            }

\date{Received 9 May 2008 / Accepted 28 August 2008}

\offprints{J.~M.~Diederik~Kruijssen, e-mail: {\tt  kruijssen@astro.uu.nl}}

\abstract{To obtain an accurate description of broad-band photometric star cluster evolution, certain effects should be accounted for. {{Energy equipartition}} leads to {{mass segregation and}} the preferential loss of low-mass stars, while stellar remnants severely influence cluster mass-to-light ratios. Moreover, the stellar initial mass function and cluster metallicity affect photometry as well. Due to the continuous production of stellar remnants, their impact on cluster photometry is strongest for old systems like globular clusters. This, in combination with their low metallicities, evidence for mass segregation, and a possibly deviating stellar initial mass function, makes globular clusters interesting test cases for cluster models.}
{In this paper we describe cluster models that include the effects of {{the preferential loss of low-mass stars}}, stellar remnants, choice of initial mass function and metallicity. The photometric evolution of clusters is predicted, and the results are applied to Galactic globular clusters.}
{The cluster models presented in this paper represent an analytical description of the evolution of the underlying stellar mass function due to stellar evolution and dynamical cluster dissolution. Stellar remnants are included by using initial-remnant mass relations, while cluster photometry is computed from the Padova 1999 isochrones.}
{Our study shows that the preferential loss of low-mass stars strongly affects cluster magnitude, colour and mass-to-light ratio evolution, as it increases cluster magnitudes and strongly decreases mass-to-light ratios. The effects of stellar remnants are prominent in the evolution of cluster mass, magnitude and mass-to-light ratio, while variations of the initial mass function induce similar, but smaller changes. Metallicity plays an important role for cluster magnitude, colour and mass-to-light ratio evolution. The different effects can be clearly separated with our models. We apply the models to the Galactic globular cluster population. Its properties like the magnitude, colour and mass-to-light ratio ranges are well reproduced with our models, provided that {{the preferential loss of low-mass stars}} and stellar remnants are included. We also show that the mass-to-light ratios of clusters of similar ages and metallicities cannot be assumed to be constant for all cluster luminosities. Instead, mass-to-light ratio increases with cluster luminosity and mass.}
{These models underline the importance of more detailed cluster models when considering cluster photometry. By including the preferential loss of low-mass stars and the presence of stellar remnants, the magnitude, colour and mass-to-light ratio ranges of modelled globular clusters are significantly altered. With the analytic framework provided in this paper, observed cluster properties can be interpreted in a more complete perspective.}

\keywords{
Galaxy: globular clusters: general --
Galaxy: open clusters and associations: general --
galaxies: star clusters --
galaxies: stellar content
}

\authorrunning{J.~M.~D.~Kruijssen and H.~J.~G.~L.~M.~Lamers}
\titlerunning{The photometric evolution of star clusters}

\maketitle


\section{Introduction} \label{sec:intro}
In recent studies, the photometric evolution of star clusters has been extensively treated from various approaches \citep[e.g.,][]{andersfritze03,lamers06,vonhippel06,fagiolini07}. Because cluster photometry is used for a broad range of applications, like age-dating galaxies and tracking their formation history, it is crucial to obtain an accurate description of the photometric evolution of clusters. While {\it Simple Stellar Population} (SSP) models \citep[e.g.,][]{leitherer99,bruzual03,andersfritze03,maraston05} only consider the changing photometric properties due to stellar evolution, other models that also use the dynamical input of $N$-body simulations can predict the photometric evolution of clusters under a wider variety of conditions \citep[e.g.,][]{lamers06,fagiolini07,borch07}. In reality, not only stellar evolution but also the dynamical interaction of a cluster with its environment causes it to lose stars \citep[e.g.,][]{baumgardt03}. This process, which is called dissolution, occurs due to internal two-body relaxation and external effects like tidal perturbation, spiral arm passages or encouters with Giant Molecular Clouds \citep[e.g.,][]{baumgardt03,gieles06,gieles07}. It can change the shape of the stellar mass function, and will also affect photometric cluster evolution. This is the case as a cluster {evolves towards energy equipartition}, causing it to preferentially lose low-mass stars \citep[e.g.,][]{portegieszwart01,baumgardt03,hurley05}.

{The physical driving force of the preferential loss of low-mass bodies is subject to debate. On the one hand, energy equipartition between the bodies constituting a cluster increases the velocities of low-mass objects and thereby gives rise to the preferential loss of low-mass stars. On the other hand, it has been proposed that mass segregation, a phenomenon in which due to energy equipartition the more massive stars sink towards the cluster centre and low-mass objects move outwards, leads to the same effect since bodies in the cluster outskirts are more loosely bound than objects in the cluster centre and are thus more easily lost \citep[e.g.,][]{leon00,portegieszwart01,lamers06}. This line of reasoning is not compatible with \citet{king66}, where it is shown that the escape rate of stars from a cluster does not vary with radius. In that scenario, the preferential loss of low-mass stars and mass segregation are both the effects of energy equipartition, but do not necessarily share any causal connection \citep[e.g.,][]{delafuentemarcos00}. Regardless of its specific nature, in the remainder of this paper we consider energy equipartition to be the fundamental cause of the preferential loss of low-mass stars: either directly, via mass segregation, or a combination of the two. In any case, mass segregation can serve as an indicator for clusters that have undergone a strong preferential loss of low-mass stars \citep[e.g.,][]{portegieszwart01,baumgardt03} and will therefore be used in that respect.}

There have been many observations of Galactic open and globular clusters in which evidence of mass segregation was found \citep[e.g.,][]{anderson96,hillenbrand98,zoccali98,albrow02,richer04,koch04,pasquali04}. These clusters can be expected to exhibit non-canonical photometric evolution. Furthermore, for a number of clusters overall mass-to-light ratios are observed that strongly deviate from the mean value presented in \citet{mclaughlin00} \citep[e.g.,][]{baumgardt03b,vandeven06}, which suggests a range of scenarios for photometric cluster evolution. This, in combination with the high mass-to-light ratio {\it in the centre} of some globular clusters \citep[e.g.,][]{pasquali04,vandenbosch06} and the consequent invocation of intermediate mass black holes (IMBHs) \citep[e.g.,][]{portegieszwart02,gurkan04}, asks for a cluster model that can explain the observed range of mass-to-light ratios. While some studies suggest IMBHs to explain the high mass-to-light ratio in the centres of globular clusters \citep[e.g.,][]{gebhardt05,noyola06}, others show that these are not required and central concentrations of stellar remnants also provide a solution \citep[e.g.,][]{baumgardt03a,baumgardt03b,hurley07}. Therefore, it is important to investigate to what extent either model can be used to explain the observed range of mass-to-light ratios, colours and magnitudes. A model describing cluster mass-to-light ratios may also be able to provide insight in the connection between globular clusters and ultra-compact dwarf galaxies (UCDs), the latter having a different mass-to-light ratio range than globular clusters \citep[e.g.,][]{hasegan05,evstigneeva07,rejkuba07,mieske08a}.

In this paper we present cluster models that are based on stellar isochrones like all SSP models, but analytically incorporate dynamical effects on cluster photometry by following the results from $N$-body simulations \citep{baumgardt03}. The resulting speed and applicability to a large parameter space makes it very suitable for studying the effect of a range of parameters on cluster evolution. We show how photometric properties of clusters like their magnitude, colour and overall mass-to-light ratio are affected by {the preferential loss of low-mass stars}, the inclusion of stellar remnants, the stellar initial mass function (IMF) and metallicity.

The structure of the paper is as follows. Our cluster evolution models are presented in Sect.~\ref{sec:clevo}. In Sect.~\ref{sec:evo} it is shown how stellar evolution affects the cluster content, including the production of stellar remnants. We derive the equations describing dynamical effects of cluster evolution on a multi-component powerlaw IMF \citep[e.g.,][]{kroupa01} in Sect.~\ref{sec:diss}. The effects of {the preferential loss of low-mass stars} and the dynamical loss of remnants are included. In that section, we also provide the final set of equations to describe cluster evolution with our models. The computation of photometric evolution is treated in Sect.~\ref{sec:phot}. The results are presented in Sect.~\ref{sec:results}, where also the influences of {the preferential loss of low-mass stars}, stellar remnants, IMF and metallicity are investigated, and the results are compared to previous studies. In Sect~\ref{sec:appl}, the models are applied to Galactic globular clusters. Section~\ref{sec:disc} contains a discussion of the results, while our conclusions are provided in Sect.~\ref{sec:concl}.

\section{Cluster evolution} \label{sec:clevo}
In this section we describe cluster evolution: first we formulate the stellar mass function in the cluster, secondly we describe cluster mass loss due to stellar evolution, and thirdly cluster mass loss due to dissolution. We include the effects of {the preferential loss of low-mass stars} and stellar remnants. {The models presented here represent the cluster evolution part of our new cluster population synthesis code {\it SPACE}\footnote{This is an acronym for {S}tellar {P}opulation {A}ge {C}omputing {E}nvironment.}, which is a fast code to predict observables like age, mass and magnitude distributions of clusters for a range of galactic conditions.}

\subsection{The stellar mass function} \label{sec:mf}
Stars in a cluster are distributed according to a mass function, for which we assume a general expression for a multi-component powerlaw mass function, which is a function of time:
\begin{equation}
\label{eq:smf}
  N_{\rm s}(t,\ms) {\rm d}\ms = C(t)\eta(\ms)\ms^{-\beta(m_{\rm s})}\d \ms ,
\end{equation}
for $\mmin < \ms < \mmax$, where $N_{\rm s}(t, \ms)$ represents the number of stars per $\msun$ at age $t$, $C(t)$ is the (time-dependent) normalisation of the mass function, $\eta(\ms)$ is introduced to preserve continuity at the stellar mass where the slope $\beta(\ms)$ changes, and $\ms$ denotes stellar mass in $\msun$. Setting $t=0$ gives the {\it initial} mass function (IMF): $N_{\rm s}(0,\ms)$. For a Salpeter IMF, we have constant values $\eta(\ms)=1$ and $\beta(\ms)=2.35$ {with $\mmini\approx 0.1~\msun$, while a Kroupa IMF} is described by
\begin{equation}
\label{eq:kroupa2}
  \beta(\ms) = \left\{
  \begin{array}{ccl}
  \beta_1 = 1.3 & {\rm for} & 0.08~\msun \leq \ms < 0.50~\msun, \\
  \beta_2 = 2.3 & {\rm for} & 0.50~\msun \leq \ms , \\
  \end{array}\right.
\end{equation}
with initial minumum stellar mass $\mmini=0.08~\msun$ and $\eta(\ms\geq 0.50~\msun)=2\eta(\ms< 0.50~\msun)$.

After cluster formation, clusters {\it lose} mass due to stellar evolution and dissolution\footnote{As shown by \citet{mieske07}, the capture of stars by clusters is ineffective.}. This section provides a description of how both mechanisms affect the stellar mass function and cluster content. Clusters with and without {the preferential loss of low-mass stars} are treated.

\subsection{Stellar evolution} \label{sec:evo}
The total mass loss due to stellar evolution and dissolution can be written as
\begin{equation}
\label{eq:dmdt}
  \frac{\d \mctot}{\d t} = \left(\frac{\d \mctot}{\d t}\right)_{\rm ev} + \left(\frac{\d \mctot}{\d t}\right)_{\rm dis} .
\end{equation}
In this section we describe the mass loss due to stellar evolution. We separate cluster mass into its two constituents: the mass in stars, or the luminous mass $\mclt$, and the mass in stellar remnants $\msrt$. {The total mass $\mctott$ is then obtained from}
\begin{equation}
\label{eq:mctot}
  \mctott = \mclt+\msrt .
\end{equation}

\subsubsection{Luminous mass}
The first right-hand term of Eq.~\ref{eq:dmdt} can be expressed in terms of a mass fraction $\muevtott\equiv\mctott/\mci$ that is still present in the cluster if stellar evolution were the only mass loss mechanism\footnote{If cluster mass also decreases due to dissolution, then the mass loss would be larger, see Sect.~\ref{sec:diss}.}:
\begin{equation}
\label{eq:dmevdt}
  \left(\frac{\d \mctot}{\d t}\right)_{\rm ev} = \frac{\d\muevtott}{\d t}\mci\frac{C(t)}{C(0)} ,
\end{equation}
where the fraction $C(t)/C(0)$ is included to correct the mass loss for a possible changing mass function normalisation due to dissolution (which will be described in Sect.~\ref{sec:diss}). Consequently, the mass fraction lost due to stellar evolution is defined by $\qevtott\equiv 1-\muevtott$. The remaining total mass fraction $\muevtott$ is the sum of the luminous mass fraction $\muevt$ and the stellar remnant (sr) mass fraction $\muevsrt$:
\begin{equation}
\label{eq:muevtot}
  \muevtott=\muevt+\muevsrt .
\end{equation}
\begin{table*}[tb]\centering
\begin{tabular}{|c c|c c c|c c c|}
  \hline \multicolumn{8}{|c|}{Luminous cluster mass ($\qev$)} \\
  \hline \hline & & & Salpeter IMF & & & Kroupa IMF & \\ \hline \hline
  Z & \mmaxi & $a_{\rm ev}$ & $b_{\rm ev}$ & $c_{\rm ev}$ & $a_{\rm ev}$ & $b_{\rm ev}$ & $c_{\rm ev}$ \\ \hline
  0.0004 & 68.5101 & 6.83 & 0.316 & -1.824 & 6.84 & 0.298 & -1.648 \\ 
  0.0040 & 69.8779 & 6.80 & 0.313 & -1.844 & 6.80 & 0.295 & -1.667 \\
  0.0080 & 71.6802 & 6.76 & 0.309 & -1.853 & 6.77 & 0.291 & -1.674 \\
  0.0200 & 68.1211 & 6.70 & 0.308 & -1.872 & 6.71 & 0.290 & -1.691 \\
  0.0500 & 49.4481 & 6.63 & 0.314 & -1.897 & 6.65 & 0.297 & -1.714 \\\hline
  \hline \multicolumn{8}{|c|}{Total cluster mass ($\qevtot$)} \\ 
  \hline \hline & & & Salpeter IMF & & & Kroupa IMF & \\ \hline \hline
  Z & \mmaxi & $a_{\rm ev}$ & $b_{\rm ev}$ & $c_{\rm ev}$ & $a_{\rm ev}$ & $b_{\rm ev}$ & $c_{\rm ev}$ \\ \hline
  0.0004 & 68.5101 & 6.93 & 0.271 & -1.855 & 6.93 & 0.255 & -1.682 \\ 
  0.0040 & 69.8779 & 6.89 & 0.271 & -1.872 & 6.90 & 0.256 & -1.696 \\
  0.0080 & 71.6802 & 6.88 & 0.265 & -1.877 & 6.88 & 0.250 & -1.701 \\
  0.0200 & 68.1211 & 6.82 & 0.263 & -1.893 & 6.83 & 0.248 & -1.716 \\
  0.0500 & 49.4481 & 6.78 & 0.263 & -1.908 & 6.79 & 0.249 & -1.731 \\\hline
\end{tabular}
\caption[]{\label{tab:qfit}
      {\it Top:} Fitting values for $a_{\rm ev}$, $b_{\rm ev}$ and $c_{\rm ev}$ for the {{\it luminous} fractional cluster mass decrease due to stellar evolution} $\qevt$ using the Padova 1999 isochrones (see text) at five different metallicities for Salpeter (0.1~$\msun<\ms<\mmaxi$) and Kroupa (0.08~$\msun<\ms<\mmaxi$) IMFs. The values of the initial maximum stellar mass $\mmaxi$ correspond to the maximum masses at the youngest isochrones ($\log{t}=6.6$). {The maximum difference between the fits and the exact $\qevt$ curves is 3\% after 19~Gyr.} {\it Bottom:} {Same as above, but for {\it total} fractional cluster mass decrease due to stellar evolution, accounting for stellar remnants. The maximum difference between the fits and the exact $\qevtott$ curves is less than 10\% after 19~Gyr.}
    }
\end{table*}
\citet{lamers05} have shown that for luminous cluster mass, $\qevt\equiv 1-\muevt$ can be approximated as
\begin{equation}
\label{eq:qev}
  \log{\qev}=(\log{t}-a_{\rm ev})^{b_{\rm ev}}+c_{\rm ev} ,
\end{equation}
for $\log{t}>a_{\rm ev}$, and where $a_{\rm ev}$, $b_{\rm ev}$ and $c_{\rm ev}$ are constants determined by the IMF and metallicity of the cluster. We use the Padova 1999 isochrones (Bertelli et al. 1994, AGB treatment as in Girardi et al. 2000) to determine the maximum stellar mass $\mmax$ that is still present in the cluster at time $t$. This also provides the initial maximum stellar mass $\mmaxi$, which is taken to be the maximum mass at the youngest isochrone ($\log t=6.6$). After assuming an IMF, i.e., $N_{\rm s}(0,\ms)$, we can write the remaining mass fraction of luminous mass as
\begin{equation}
\label{eq:muev}
  \muevt= \frac{\int_{m_{\rm min,i}}^{m_{\rm max}(t)}\ms N_{\rm s}(0,\ms)\d \ms}{\mci} ,
\end{equation}
from which exact values for $\qevt\equiv 1-\muevt$ can be determined. The resulting fitting constants $a_{\rm ev}$, $b_{\rm ev}$ and $c_{\rm ev}$ are summarised in the top half of Table~\ref{tab:qfit}\footnote{Values for the case of a Salpeter IMF are also provided by \citet{lamers05}, but these are based on an older version of the $GALEV$ models \citep{schulz02,andersfritze03}.}. The method assumes the instantaneous removal of stars and ignores mass loss by stellar winds, but this assumption is legitimate since massive stars hardly contribute to overall cluster mass and low-mass stars only suffer significant mass loss during the last 10\% of their lifetime. In order to determine $\muevtott$ and evaluate Eq.~\ref{eq:dmevdt} also $\muevsrt$ is required, which will be discussed separately in Sect.~\ref{sec:remnprod}.

Since in the present study {the stellar content of clusters is considered}, we will be evolving the stellar mass function of a cluster (and thereby indirectly the cluster mass) rather than evaluating cluster mass itself. Therefore, we will use $\mmax$ from the isochrones instead of Eq.~\ref{eq:dmevdt} to incorporate stellar evolution. The time evolution of $\mmax$ is shown in the left-hand panel of Fig.~\ref{fig:mremn} for metallicities $Z=\{0.0004,0.004,0.02\}$.

\subsubsection{Stellar remnants} \label{sec:remnprod}
Stellar remnants can constitute a large fraction of the total cluster mass\footnote{{If stellar mass-dependent cluster dissolution is omitted, after 12~Gyr stellar remnants constitute about 30\% of the total cluster mass. Within this fraction, the summed mass ratios of black holes, neutron stars and white dwarfs are 1:3.5:9. These ratios change when mass-dependent dissolution is included.}}. Especially during the later stages of cluster lifetime, the cluster content is likely to be dominated by remnants. As will be shown in Sect.~\ref{sec:results} several observables can be significantly affected, making it essential to include remnants when studying cluster evolution. 

The distinction between total and luminous cluster mass can be made by writing $\mctot=\mcl+\msr$, where $\msr$ denotes the part of the total cluster mass constituted by stellar remnants. The evolution of dark cluster mass can be expressed as
\begin{equation}
\label{eq:dmsrdt}
  \frac{\d \msr}{\d t} = \left(\frac{\d \msr}{\d t}\right)_{\rm ev} + \left(\frac{\d \msr}{\d t}\right)_{\rm dis} ,
\end{equation}
where at the right-hand side the first (positive) term describes the production of remnants due to stellar evolution, and the second (negative) term represents remnant loss due to dissolution. Stellar evolution removes stars at the high-mass end of the mass function. Stellar remnant production can be included by assuming an initial-remnant mass relation and leaving a remnant mass $\mssr$ upon such removal.
\begin{figure*}[t]
\resizebox{\hsize}{!}{\includegraphics{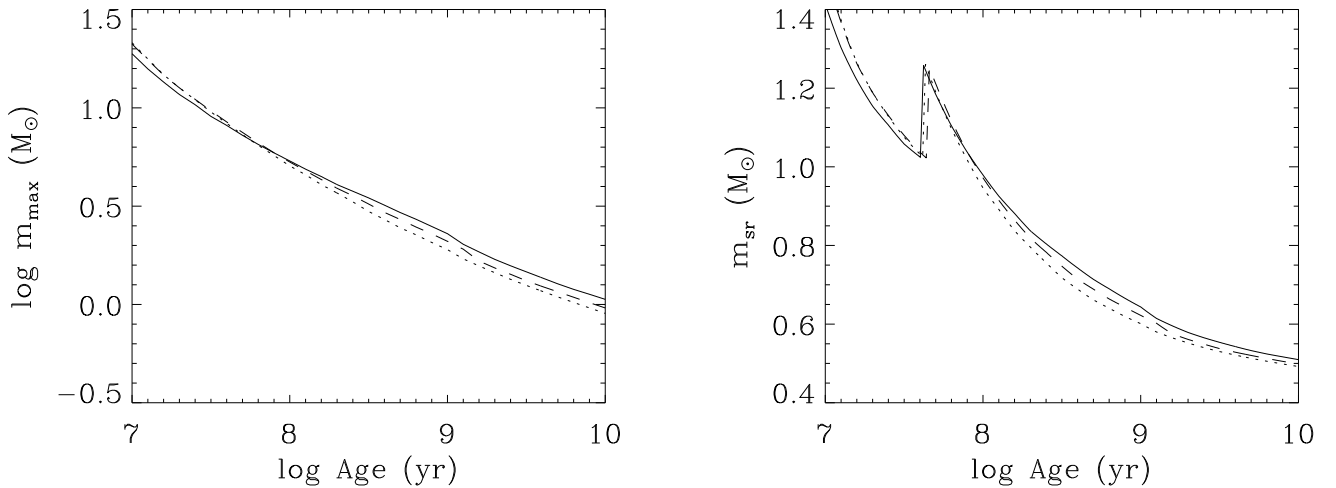}}
\caption[]{\label{fig:mremn}
      {\it Left}: Maximum stellar mass $\mmax$ as a function of age for $Z=\{0.0004,0.004,0.02\}$ (dotted, dashed and solid lines, respectively). {\it Right:} Produced stellar remnant mass $\mssrt$ as a function of age for $Z=\{0.0004,0.004,0.02\}$ (dotted, dashed and solid lines, respectively). The peak corresponds to the transition from neutron star production to white dwarf production, i.e., the lifetime of an $8~\msun$ star.
    }
\end{figure*}

From \citet{kalirai07}, for white dwarfs ($\ms<8~\msun$) this relation is given by
\begin{equation}
\label{eq:wd}
  \mssr = 0.109~\ms+0.394~\msun,
\end{equation}
with $\mssr$ the remnant mass and $\ms$ the stellar mass, where $\ms$ is equal to the initial stellar mass in our models since instantaneous death is assumed. For $\ms<0.45$~\msun, Eq.~\ref{eq:wd} is not valid since the remnant mass then exceeds the progenitor mass. However, this does not cause any problems because it only occurs for ages {\it much} larger than the Hubble time. 

For neutron stars ($8~\msun\leq \ms<30~\msun$), the initial-remnant mass relation is taken to be \citep{nomoto88}:
\begin{equation}
\label{eq:ns}
  \mssr = 3.63636\times 10^{-2}~(\ms-8~\msun)+1.02~\msun .
\end{equation}

Stars with masses $\ms\geq 30~\msun$ result in a black hole and since no definitive models for black hole formation exist, the corresponding remnant mass is assumed to be
\begin{equation}
\label{eq:bh}
  \mssr = 8~\msun .
\end{equation}
This value is in agreement with dynamical masses obtained from X-ray binary observations \citep[e.g.,][]{casares06}. When the above initial-remnant mass relation is linked to the maximum stellar mass $\mmax$, the remnant mass $\mssrt$ that is produced at time $t$ can be obtained. This varies with metallicity; for $Z=\{0.0004,0.004,0.02\}$ examples are shown in the right-hand panel of Fig.~\ref{fig:mremn}.

We can now compute the total cluster mass fraction in stellar remnants $\muevsrt$ by integrating the remnant masses $\mssr$ over the mass function for the mass range of all stars that have ended their lives:
\begin{equation}
\label{eq:muevsr}
  \muevsrt= \frac{\int_{m_{\rm max}(t)}^{m_{\rm max,i}}\mssr(\ms) N_{\rm s}(0,\ms)\d \ms}{\mci} ,
\end{equation}
with $N_{\rm s}(0,\ms)$ the initial form of Eq.~\ref{eq:smf}. The total remaining mass fraction $\muevtott$ is obtained by adding this to the mass fraction in luminous mass (see Eq.~\ref{eq:muevtot}). {Similarly to $\qev$, we provide fitting constants as in Eq.~\ref{eq:qev} for $\qevtot\equiv 1-\muevtot$ in the bottom half of Table~\ref{tab:qfit}. Again, we will not use these fits in the remainder of this study, since the stellar content of clusters is treated. Instead, the initial-remnant relations from Eqs.~\ref{eq:wd}---\ref{eq:bh} are used. Nonetheless, Table~\ref{tab:qfit} is provided for convenience.}

Given the expression for $\muevsrt$, the remnant production per unit time (first right-hand term of Eq.~\ref{eq:dmsrdt}) can now be determined. By multiplying the time derivative of $\muevsrt$ with initial cluster mass and correcting the normalisation of the stellar mass function for a possible decrease due to dissolution, this fraction can be translated into the time derivative of non-luminous cluster mass
\begin{equation}
\label{eq:dmsrdtev}
  \left(\frac{\d \msr}{\d t}\right)_{\rm ev} = \frac{\d\muevsrt}{\d t}\mci\frac{C(t)}{C(0)} ,
\end{equation}
which is similar to the expression derived for the total cluster mass evolution due to stellar evolution (Eq.~\ref{eq:dmevdt}). 

{The above approach does not account for kick velocities that are obtained by neutron stars and black holes upon their creation. The implications for integrated cluster properties are discussed in Sect.~\ref{sec:assump}, point (3).}

\subsection{Dissolution} \label{sec:diss}
The effect of dissolution on the stellar mass function differs for clusters with or without {the preferential loss of low-mass stars}. In clusters {that have reached energy equipartition}, the low-mass objects tend to reside in the outer regions, while the most massive bodies are generally located near the cluster centre, and the cluster preferentially loses low-mass objects. In clusters without {the preferential loss of low-mass stars}, bodies\footnote{`Bodies' denotes either stars or stellar remnants.} of different masses are lost with almost equal probabilities. Both cases are treated in this section.

The dynamical mass loss (second right-hand term of Eq.~\ref{eq:dmdt}) can be expressed as
\begin{equation}
\label{eq:dmdtdis}
  \left(\frac{\d \mctot}{\d t}\right)_{\rm dis} = -\frac{\mctot}{\tau_{\rm dis}} = -\frac{\mctot}{t_0\mctot^\gamma} = -\frac{\mctot^{1-\gamma}}{t_0} ,
\end{equation}
where $\tau_{\rm dis}$ is the dissolution timescale. In empirical studies it was shown to be related to present cluster mass as $\tau_{\rm dis}=t_0(\mc/\msun)^\gamma$ by \citet{boutloukos03} and \citet{lamers05}. \citet{boutloukos03} found $\gamma=0.62\pm 0.06$, the same value as derived by \citet{gieles04} for the $N$-body simulations of clusters in tidal fields by \citet{baumgardt03}. Please note that $\mctot$ represents the total cluster mass and thus includes stars as well as their remnants.

{The rapidity of the exponential mass decrease in Eq.~\ref{eq:dmdtdis} is set by the dissolution timescale parameter $t_0$, which is essentially the dissolution timescale for a hypothetical $1~\msun$ cluster. Because the total disruption time $\tdis$ of a cluster is determined by dynamical dissolution {\it and} stellar evolution and thus depends on the adopted IMF and on metallicity, we prefer the use of $t_0$ over $\tdis$ to characterise the strength of dissolution alone. Throughout this paper we will give the typical total disruption times associated with used values of the dissolution timescale $t_0$.}

\subsubsection{Mass loss from clusters without {the preferential loss of low-mass stars}: the `canonical mode'} \label{sec:dissnoseg}
Cluster mass evolution can be computed numerically from its time derivatives due to stellar evolution and dissolution (Eqs.~\ref{eq:dmevdt} and~\ref{eq:dmdtdis}). For a more detailed and more accurate description of cluster evolution we turn to its stellar content. We distinguish between stellar mass-independent dissolution (`canonical mode') and the preferential loss of low mass stars (`{preferential} mode'). When considering mass loss in the canonical mode, i.e., the evolution of the stellar mass function of a cluster without {the preferential loss of low-mass stars}, the influence of mass loss due to stellar evolution and dissolution is twofold. Stellar evolution decreases the maximum stellar mass $\mmax$, while dissolution decreases the normalisation factor of the mass function $C(t)$ (see Eq.~\ref{eq:smf}) because stellar masses are randomly distributed throughout the cluster and thus all bodies have approximately similar probabilities of being ejected. 

The normalisation of the mass function $C(t)$ is directly proportional to luminous cluster mass $\mclt$ because the latter is obtained by integrating the mass function (Eq.~\ref{eq:smf}) over stellar mass. We define a parameter $f_{\rm pref}(t)$ to describe the fraction of mass loss occurring in {preferential} mode (see Sect.~\ref{sec:dissseg}) and a fraction $f_{\rm sr}(t)$ of the mass loss to occur in the form of stellar remnants. This implies that the factor $1-f_{\rm pref}(t)$ denotes the fraction of mass loss that occurs in the canonical mode, while the factor $1-f_{\rm sr}(t)$ represents the fraction of mass loss that takes place in the form of luminous mass (i.e., not in stellar remnants). For mass loss in the canonical mode, i.e., without {the preferential loss of low-mass stars}, we have $f_{\rm pref}(t)=0$. Then the time derivative of luminous cluster mass from a cluster in the canonical mode, i.e., without {the preferential loss of low-mass stars}, is written as
\begin{eqnarray}
\label{eq:dc1}
  \nonumber \left(\frac{\d\mcl}{\d t}\right)_{\rm dis}^{\rm can}&=&[1-f_{\rm pref}(t)][1-f_{\rm sr}(t)]\left(\frac{\d\mctot}{\d t}\right)_{\rm dis} \\
                                                     &=&\frac{\mcl}{C(t)}\frac{\d C}{\d t} ,
\end{eqnarray}
where $C\propto\mcl$ and thus $\d\ln{C}/\d t=\d\ln{\mcl}/\d t$ leads to the last equality, {and the label `can' indicates we are dealing with mass loss in the canonical mode}. Substituting the total mass derivative from Eq.~\ref{eq:dmdtdis} yields after rearranging
\begin{equation}
\label{eq:dcdt1}
  \frac{\d C}{\d t} = -[1-f_{\rm pref}(t)][1-f_{\rm sr}(t)]\frac{\mctot^{1-\gamma}}{t_0\mcl}C(t) .
\end{equation}
In the canonical mode, the influence of mass loss on the stellar mass function is completely described by this expression. If either the cluster { only loses stars of the lowest masses ($f_{\rm pref}(t)=1$) or all mass is lost in the form of stellar remnants ($f_{\rm sr}(t)=1$), we have constant normalisation $C(t)$ and mass loss affects the cluster content in different ways (see below and Sect.~\ref{sec:dissseg})}.

To include the loss of stellar remnants, we specify the expression for the fraction of mass loss that occurs in the form of remnants $f_{\rm sr}(t)$. For mass loss in the canonical mode the spatial distribution of bodies is random. This implies that the fraction of the total mass lost due to dissolution that is lost in the form of remnants is equal to the ratio of remnant mass to total cluster mass
\begin{equation}
\label{eq:fsr}
  f_{\rm sr}(t) = \msrt/\mctott .
\end{equation}
We can use this parameter to describe how the cluster mass in stellar remnants changes due to dissolution (the second right-hand term of Eq.~\ref{eq:dmsrdt}):
\begin{eqnarray}
\label{eq:dmsrdtdis}
  \nonumber \left(\frac{\d \msr}{\d t}\right)_{\rm dis}^{\rm can}&=&[1-f_{\rm pref}(t)]f_{\rm sr}(t)\left(\frac{\d \mctot}{\d t}\right)_{\rm dis} \\
                                                      &=&-[1-f_{\rm pref}(t)]f_{\rm sr}(t)\frac{\mctot^{1-\gamma}}{t_0} ,
\end{eqnarray}
where we use the description of the total mass loss derivative due to dissolution from Eq.~\ref{eq:dmdtdis}. Again, the factor $1-f_{\rm pref}(t)$ represents the fraction of mass loss {that takes place} in the canonical mode, while the factor $f_{\rm sr}(t)$ denotes the fraction of mass loss that occurs in the form of stellar remnants. This expression completely describes the influence of mass loss in the canonical mode on the total cluster mass in stellar remnants.

\subsubsection{Mass loss from clusters including {the preferential loss of low-mass stars}: the `{preferential} mode'} \label{sec:dissseg}
From $N$-body simulations, it is shown by \citet{baumgardt03} how {energy equipartition} affects the evolution of the stellar mass function. {From their study it is evident that most clusters exhibit the preferential loss of low-mass stars, making it a very important mechanism when considering cluster evolution.} In \citet{lamers06}, analytical models for the evolution of the mass function are presented which are based on these simulations. In their models, dissolution no longer induces a uniform effect on the mass function by decreasing its normalisation. Instead, preferentially low-mass stars are removed. This can be approximated by a gradual increase in the lower mass limit of the stars present in the cluster $\mmin$ \citep{lamers06}. By using this description, the {\it slope} of the mass function remains unchanged as the loss of low-mass stars is incorporated by increasing the lower mass limit. This is done in such a way that the mean stellar mass is always comparable to the mean stellar mass in the $N$-body simulations by \citet{baumgardt03}, in which the shape of the mass function {\it does} change due to the preferential loss of low-mass stars. 

To describe mass loss in the {preferential} mode, i.e., mass loss including {the preferential loss of low-mass stars}, we use the parameter $f_{\rm pref}(t)$ that represents the fraction of cluster mass loss that occurs in {preferential} mode at time $t$. The formulation allows for intermediate cases of {energy equipartition}, since $f_{\rm pref}(t)$ can have any value between 0 and 1. The $N$-body simulations by \citet{baumgardt03} suggest a rapid transition from a random ejection of bodies to {the preferential loss of low-mass stars}, resulting in almost a step function for $f_{\rm pref}(t)$. For a cluster {that initially does not preferentially lose low-mass stars} but {reaches complete energy equipartition} at $t=t_{\rm pref}$ we can then write
\begin{equation}
\label{eq:fseg}
  f_{\rm pref}(t) = \left\{
  \begin{array}{ccl}
  0 & {\rm for} & t < t_{\rm pref}  \\
  1 & {\rm for} & t \geq t_{\rm pref} . \\
  \end{array}\right.
\end{equation}
However, because complete {energy equipartition} ($f_{\rm pref}(t)=1$) is unlikely to occur (see Sect.~\ref{sec:results}), we will probably have $f_{\rm pref}(t)<1$ for $t \geq t_{\rm pref}$.

By explicitly integrating stellar mass over the mass function (Eq.~\ref{eq:smf}), the luminous cluster mass is obtained. Because in the {preferential} mode dissolution only affects the minimum stellar mass, we can write for the time derivative of luminous cluster mass:
\begin{eqnarray}
\label{eq:dmmin1}
  \nonumber \left(\frac{\d\mcl}{\d t}\right)_{\rm dis}^{\rm pref}&=&f_{\rm pref}(t)[1-f_{\rm sr}^{\rm pref}(t)]\left(\frac{\d\mctot}{\d t}\right)_{\rm dis} \\
                                                     &=&-\frac{C(t)\eta(m_{\rm min})}{2-\beta(m_{\rm min})}\frac{\d}{\d t}m_{\rm min}^{2-\beta(m_{\rm min})} ,
\end{eqnarray}
for $\beta(m_{\rm min})\neq 2$ and
\begin{eqnarray}
\label{eq:dmmin2}
  \nonumber \left(\frac{\d\mcl}{\d t}\right)_{\rm dis}^{\rm pref}&=&f_{\rm pref}(t)[1-f_{\rm sr}^{\rm pref}(t)]\left(\frac{\d\mctot}{\d t}\right)_{\rm dis} \\
                                                     &=&-C(t)\eta(m_{\rm min})\frac{\d}{\d t}\ln(m_{\rm min}) ,
\end{eqnarray}
if $\beta(m_{\rm min})=2$, where $\beta(m_{\rm min})$ is the slope of the stellar mass function (see Eq.~\ref{eq:smf}) {and the label `pref' indicates we are dealing with mass loss in the preferential mode}. The fraction of mass loss occurring in the {preferential} mode is represented by the factor $f_{\rm pref}(t)$, while the factor $1-f_{\rm sr}^{\rm pref}(t)$ denotes the fraction of mass loss {in the preferential mode} occurring in the form of luminous mass (i.e., not in the form of remnants, see below for more details). Again, we substitute the total mass derivative (Eq.~\ref{eq:dmdtdis}), which yields a differential equation describing the effect of the preferential loss of low-mass stars. After rearranging the terms, mass loss in the {preferential} mode thus results in an evolution of the lower mass limit $\mmin$:
\begin{equation}
\label{eq:dmmin}
\frac{\d m_{\rm min}}{\d t}=f_{\rm pref}(t)[1-f_{\rm sr}^{\rm pref}(t)]\frac{\mctot^{1-\gamma}m_{\rm min}(t)^{\beta(m_{\rm min})-1}}{C(t)\eta(m_{\rm min})t_0} ,
\end{equation}
for all values of $\beta(m_{\rm min})$. If a cluster is {has not reached energy equipartition} ($f_{\rm pref}(t)=0$), mass loss is independent of stellar mass. In that case, the time derivative $\d m_{\rm min}/\d t=0$ and we thus have constant $\mmin$.

We now describe stellar remnant loss from clusters {that have reached energy equipartition}. If a cluster is completely mass-segregated, remnants are produced in the cluster centre and only become available for dissolution if remnants are the least massive bodies in the cluster. Therefore, the fraction of mass loss taking place in the form of stellar remnants that is used here $f_{\rm sr}^{\rm pref}(t)$ differs from the expression for canonical mass loss (Eq.~\ref{eq:fsr}).

From Fig.~\ref{fig:mremn} we know that the produced remnant mass nearly always decreases with time, while $\mmin$ is a monotonously increasing function of $t$. This implies that there is a time $t_{\rm sr}$ at which $\mmin$ increases to a value larger than the smallest stellar remnant mass $\mssrt$\footnote{Incidentally, because the produced stellar remnant mass nearly always descreases with time, the smallest remnant mass at time $t$ is the stellar remnant mass that is produced at that time.}. For all $t\geq t_{\rm sr}$ mass can be lost in the form of remnants\footnote{Strictly spoken, this merely holds for $\mmin$ that intersect $\mssrt$ only once. From Fig.~\ref{fig:mremn} we see that at the transition from neutron star production to white dwarf production there is a possibility for curves of $\mmin$ to increase to a value larger than $\mssrt$, before briefly being overtaken again due to the change of produced remnant type. In that case, $\mmin$ intersects $\mssrt$ three times. However, the slope of $\mmin$ is typically very steep compared to $\mssrt$, and thus only a very small fraction of all cluster initial masses will pass through the described fluctuation over a negligible timespan. Therefore, we can indeed assume that for all $t\geq t_{\rm sr}$ mass can be lost in the form of stellar remnants.}. For these values of $t$, all remnant masses $\mssrt\leq \mssr\leq\mmin$ are considered to be immediately available for dissolution (see Appendix~\ref{sec:app} for a justification of this assumption). Consequently, they are immediately {\it lost} by dissolution, because they are the least massive bodies in the cluster.

The fraction of mass loss occurring in the form of stellar remnants is different for clusters with and without {the preferential loss of low-mass stars}. Therefore, we define a separate parameter $f_{\rm sr}^{\rm pref}(t)$ for the fraction of mass loss in the form of remnants if the cluster {preferentially loses low-mass bodies}. The time derivative of dark cluster mass (the second right-hand term in Eq.~\ref{eq:dmsrdt}) thus becomes:
\begin{eqnarray}
\label{eq:dmsrdtdisseg}
 \nonumber \left(\frac{\d \msr}{\d t}\right)_{\rm dis}^{\rm pref}&=&f_{\rm pref}(t)f_{\rm sr}^{\rm pref}(t)\left(\frac{\d \mctot}{\d t}\right)_{\rm dis} \\
                                                     &=&-f_{\rm pref}(t)f_{\rm sr}^{\rm pref}(t)\frac{\mctot^{1-\gamma}}{t_0} ,
\end{eqnarray}
which is nearly the same expression as Eq.~\ref{eq:dmsrdtdis} that is valid in the canonical mode. The fraction of mass loss occurring in the {preferential} mode is again represented by the factor $f_{\rm pref}(t)$. We have yet to specify the fraction of mass loss in the form of remnants for a {the preferential loss of low-mass stars} $f_{\rm sr}^{\rm pref}(t)$. If we consider a certain time interval $\d t$, there are two possibilities: either the remnant mass available for dissolution ($M_{\rm cl}^{\rm sr,dis}$, which is the total mass of all remnants with masses smaller than the lowest stellar mass present at time $t$) is so large that all mass loss during the interval $\d t$ can be accounted for by removing remnants {\it only}, or luminous mass has to be lost as well. In the former case, the minimum stellar mass $\mmin$ is not reached before the end of the time interval $\d t$ while losing the lowest mass bodies (i.e., remnants) and all mass loss takes place in the form of stellar remnants. For the latter case, $\mmin$ {\it is} reached during the interval $\d t$, and the fraction of mass loss to occur in the form of remnants is then described by the ratio of the available remnant mass and the total mass loss during the time interval $\d t$. This implies
\begin{equation}
\label{eq:fsrseg}
  f_{\rm sr}^{\rm pref}(t) = {\rm min}\left[1,-M_{\rm cl}^{\rm sr,dis}\Big{/}\frac{\d\mctot}{\d t}\d t\right] ,
\end{equation}
where the denominator of the second term between brackets can be obtained from the expression for cluster mass loss due to dissolution (Eq.~\ref{eq:dmdtdis}) and the numerator is computed numerically from
\begin{eqnarray}
\label{eq:mclsrav}
  \nonumber M_{\rm cl}^{\rm sr,dis}(t) &=& \int_{m_{\rm max}(t)}^{m_{\rm s}^{\rm min,sr}(t)}\mssr(\ms) N_{\rm s}(t,\ms)\d \ms \\
               && -\int_{0}^{t}f_{\rm pref}(t')f_{\rm sr}^{\rm pref}(t')\frac{\mbox{$M_{\rm cl}^{\rm tot}(t')^{1-\gamma}$}}{\mbox{$t_0$}}\d t' .
\end{eqnarray}
The first right-hand term denotes the total produced remnant mass that is available for a given {\it present} mass function $N_{\rm s}(t,\ms)$ and increases with time. The upper integration limit of the integral $\ms^{\rm min,sr}(t)$ represents the initial stellar mass corresponding to remnants with mass equal to $\mmin$\footnote{These are the most massive remnants that are available for dissolution.} and the mass function is thus integrated for produced remnant masses $\mssrt\leq\mmin$. By introducing the second right-hand term in Eq.~\ref{eq:mclsrav}, we subtract the part of the produced remnant mass that has already been lost by dissolution. This integral follows from the time derivative of non-luminous cluster mass (Eq.~\ref{eq:dmsrdtdisseg}) and describes all remnant mass that was lost in the {preferential} mode. Using the present mass function in the first right-hand term of Eq.~\ref{eq:mclsrav} assumes that any change in the normalisation of the stellar mass function has a proportional effect on total remnant mass. The normalisation only changes for mass loss in the canonical mode (i.e., mass loss from clusters without {the preferential loss of low-mass stars}), for which dissolution is stellar mass-independent. Hence, if the normalisation of the stellar mass function were to change, the total mass in remnants would be affected accordingly, thereby justifying the above assumption.

\subsection{Total cluster evolution} \label{sec:total}
The description of cluster evolution from Sects.~\ref{sec:mf} to~\ref{sec:diss} includes stellar evolution and four modes of mass loss: luminous mass loss and stellar remnant loss from clusters without {the preferential loss of low-mass stars}, and luminous mass loss and stellar remnant loss from clusters {that have reached energy equipartition and do preferentially lose low-mass stars}. In this section the derived expressions are combined.

\begin{figure*}[t]
\resizebox{\hsize}{!}{\includegraphics{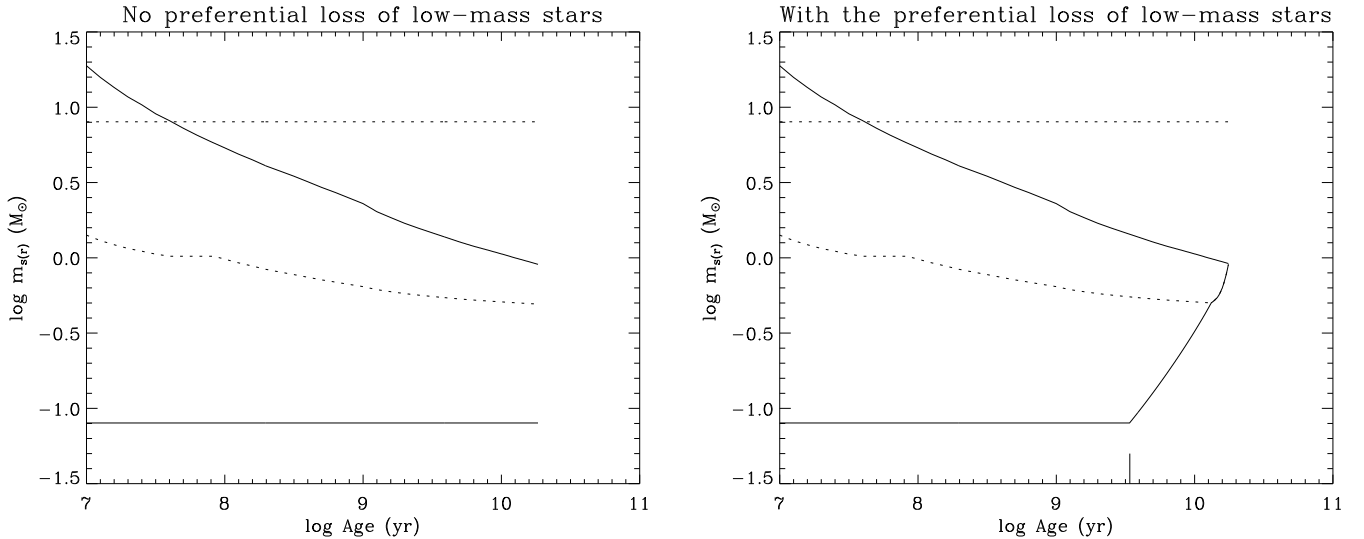}}
\caption[]{\label{fig:mrange}
      {\it Left}: Evolution of minimum and maximum masses of stars (solid) and stellar remnants (dotted) in a cluster with $\mcli=10^6~\msun$, $t_0=3$~Myr and without {the preferential loss of low-mass stars}. {\it Right:} Same as left, but including {the preferential loss of low-mass stars}. The tickmark at $\log{t}\approx 9.5$ indicates the onset of {the preferential mode (see Sect.~\ref{sec:fxseg})}.
          }
\end{figure*}
The evolution of the stellar mass function is completely described by (1) the time derivative of its normalisation factor $C(t)$ (Eq.~\ref{eq:dcdt1}) that describes the general decrease of the mass function, (2) the time derivative of the minimum stellar mass $\mmin$ (Eq.~\ref{eq:dmmin}) that describes the loss of low-mass stars, and (3) $\mmax$ that describes the loss of high-mass stars due to stellar evolution. When combining these equations, the luminous cluster mass follows from
\begin{equation}
\label{eq:mclum}
  \mclt = \int_{m_{\rm min}(t)}^{m_{\rm max}(t)}\ms N_{\rm s}(t,\ms)\d \ms ,
\end{equation}
with $N_{\rm s}(t,\ms)$ the stellar mass function from Eq.~\ref{eq:smf}.

Similarly, the expressions describing stellar remnant loss for clusters with and without {the preferential loss of low-mass stars} (Eqs.~\ref{eq:dmsrdtdis} and~\ref{eq:dmsrdtdisseg}) can be combined with the expression representing remnant production (Eq.~\ref{eq:dmsrdtev}) to write
\begin{eqnarray}
\label{eq:dmsrdt!}
  \frac{\mbox{$\d$} \msr}{\mbox{$\d t$}} &=& \frac{\mbox{$\d$}\muevsrt}{\mbox{$\d t$}}\mci\frac{\mbox{$C(t)$}}{\mbox{$C(0)$}} \\
  \nonumber &-&\left\{[1-f_{\rm pref}(t)]f_{\rm sr}(t)+f_{\rm pref}(t)f_{\rm sr}^{\rm pref}(t)\right\}\frac{\mbox{$\mctot^{1-\gamma}$}}{\mbox{$t_0$}} ,
\end{eqnarray}
with $\muevsrt$ from Eq.~\ref{eq:muevsr}. The first right-hand term denotes the production of stellar remnants, while the second represents the loss of stellar remnants in the canonical mode (represented by the factor $[1-f_{\rm pref}(t)]f_{\rm sr}(t)$) and in the {preferential} mode (represented by the factor $f_{\rm pref}(t)f_{\rm sr}^{\rm pref}(t)$). This equation can be integrated for the total remnant mass in the cluster $\msrt$. 
Finally, addition of the total remnant mass to the luminous cluster mass from Eq.~\ref{eq:mclum} yields the total cluster mass {as formulated in Eq.~\ref{eq:mctot}.}

The above set of equations can be solved numerically and represents a range of models that describe the complete evolution of cluster content for clusters with and without {the preferential loss of low-mass stars}. A simple recursive integration scheme is used. As a criterion for total cluster disruption, we use a lower luminous cluster mass limit $\mcl=100~\msun$, though other values can be adopted if necessary.

We now briefly illustrate the time evolution of the mass ranges of stars and remnants in a $10^6~\msun$ cluster. These are shown in Fig.~\ref{fig:mrange} for models with and without {the preferential loss of low-mass stars}. Solid lines describe the upper and lower mass limits of the stars in the cluster, while the minimum and maximum masses of stellar remnants are represented by dotted curves. The left panel shows the evolution without {the preferential loss of low-mass stars}, while for the right panel it is included. 

Without {the preferential loss of low-mass stars}, the minimum stellar mass is constant since mass loss occurs by removing stars of all masses and thus by decreasing the normalisation of the mass functions of stars and remnants. Stellar evolution causes the maximum stellar mass to decrease. The maximum remnant mass remains constant at the maximum remnant mass that was produced in the cluster. On the other hand, the minimum remnant mass decreases as remnants of lower masses are produced. Until total cluster disruption, bodies from a broad mass range can be retained. The cluster is completely disrupted when the normalisation constant $C(t)$ approaches zero.

When including {the preferential loss of low-mass stars}, the minimum stellar mass starts to increase as soon as mass loss occurs in the {preferential} mode. The maximum stellar and remnant masses exhibit the same behaviour as for clusters without {the preferential loss of low-mass stars}. Initially, the same holds for the minimum remnant mass. However, when the minimum remnant mass reaches the minimum stellar mass, the lowest mass bodies in the cluster are stars and remnants. This leads to a combined evolution of the lower mass limits of both. The cluster is completely disrupted when the stellar mass limits meet.

\section{Computation of photometric cluster evolution} \label{sec:phot}
Cluster photometry is computed from the Padova 1999 isochrones, that are described in \citet{bertelli94} and are based on spectral energy distributions from \citet{kurucz92}, but use a treatment of AGB stars as in \citet{girardi00}. The photometry computation is accomplished by direct integration of luminosities over the stellar mass function for a given age and initial cluster mass. This approach allows for greater flexibility when including {the preferential loss of low-mass stars}, since the evolving mass function is explicitly included in the computation. If existing SSP models had been adopted, this would not have been the case because such models only include fading by stellar evolution for a fixed mass function.

For a cluster of arbitrary age $t$ and initial mass $\mci$, the stellar luminosities of the two isochrones at ages $t_i$, $i=\{1,2\}$ closest to $t$ are integrated over the computed mass functions {\it at these ages}, with $t_1<t$ and $t_2>t$. This results in total cluster luminosities $L_{\rm cl,\lambda}(t_i,\mci)$ for passband $\lambda$. These luminosities are then interpolated to obtain $L_{\rm cl,\lambda}(t,\mci)$ and converted to absolute magnitudes. 

For existing SSP models, which only account for the effect of stellar evolution and therefore do not treat clusters near their total disruption, the above procedure suffices to determine photometric cluster evolution. However, if $t_1<t<t_{\rm dis}<t_2$ the cluster is disrupted before $t_2$ and there are no stars left at $t_2$. This leads to inadequate luminosity computations if the above interpolation is used. In that case the stellar mass function of the cluster {\it at age} $t$ is adopted for both ages $t_1$ and $t_2$ and the mass range is shifted to fit the appropriate maximum stellar masses at these ages. After integrating the resulting two mass functions, we obtain two luminosities $L_{\rm cl,\lambda}(t_i,\mci)$ for each passband. Interpolation then yields $L_{\rm cl,\lambda}(t,\mci)$. The described method does not accurately reproduce the luminosity contribution of the lowest stellar masses because the mass function is shifted to fit to $M_{\rm max}(t_1,t_2)$. However, this leaves the resulting magnitude almost completely unaffected since the high-mass end of the mass function completely dominates cluster photometry.

Contrary to some existing SSP models like {\it GALEV}, no gas emission is included in our photometric models. Line emission is only important for clusters that contain massive stars that emit ionising photons. This implies that our photometry can be considered to be accurate for ages $t\simgreat 8$~Myr for solar metallicity and $t\simgreat 20$~Myr if $Z=0.0004$ \citep{andersfritze03}.

\section{Photometric properties of clusters} \label{sec:results}
In this section we apply the models described in Sects.~\ref{sec:clevo} to~\ref{sec:phot} to investigate the effects of {the preferential loss of low-mass stars}, stellar remnants, IMF and metallicity on the mass, magnitude, colour and mass-to-light ratio evolution of clusters. We computed our models for cases with and without {the preferential loss of low-mass stars}, with and without stellar remnants, Salpeter and Kroupa IMFs, and metallicities $Z=\{0.0004, 0.004, 0.008, 0.02, 0.05\}$. Moreover, we considered initial cluster masses $\mcli$ in the range $10^2$---$10^7~\msun$ and dissolution timescales $t_0$ of 0.1---100~Myr\footnote{{This is the typical dissolution timescale range \citep[e.g.,][]{lamers05a}, corresponding to $\tdis\approx10^8$---10$^{11}$~yr for a 10$^5~\msun$ cluster. It can be easily checked by considering that the total disruption time is of the order of $\tdis\sim t_0(\mcli/\msun)^\gamma$. Throughout this section $t_0=3$~Myr will be used, which is the mean value of this range and is the typical timescale for clusters in the solar neighbourhood. This value is typical of tidally dissolving globular clusters on circular orbits at 3~kpc from the Galactic centre, or clusters on eccentric ($e=0.7$) orbits with an apogalactic distance of 8.5~kpc \citep{baumgardt03}.}}. As a result, cluster evolution for total disruption times $\tdis>10$~Myr has been computed for cluster ages between 10~Myr and 19~Gyr (the upper age limit of the stellar isochrones). {Models for a broad range of parameters are publicly available  in electronic form at the CDS and also at \texttt{http://www.astro.uu.nl/\~{}kruijs}, while predictions for specific models can be made by the first author upon request}. The most important results of our models are discussed in this section.

\subsection{The effects of model components}
{Accounting for the preferential loss of low-mass stars and including the mass of stellar remnants both have their effects in the framework of our models. In this section these effects are considered.}

\subsubsection{The effects of the preferential loss of low-mass stars} \label{sec:fxseg}
The preferential loss of low-mass stars induced by {energy equipartition and possibly also mass segregation (see Sect.~\ref{sec:intro})} can be expected to have a significant effect on the magnitude and colour evolution of clusters \citep{lamers06}, but also on their mass and mass-to-light ratio. Complete {energy equipartition} ($f_{\rm pref}(t)=1$) implies that {\it only} bodies of the lowest masses are lost, an implication that does not seem likely to be in accordance with reality for two reasons. First of all, external perturbations are not strictly confined to the very outer radius of a cluster due to internal cluster dynamics. Therefore, dynamical mass loss is not confined to the very outer layer of the cluster, which results in the loss of stars or remnants with masses above $\mmin$. Secondly, complete energy equipartition may not be reached by a cluster \citep{baumgardt03}, inducing only a partial {preferential loss of low-mass stars}. Therefore, we tuned the evolution of $\mmin$ in our complete models to the $N$-body simulations by \citet{baumgardt03} so that their mean stellar mass evolution is similar. This allows us to determine a step function form for $f_{\rm pref}(t)$ that leads to a mean mass evolution that corresponds best to its counterpart in the $N$-body simulations. This analysis results in
\begin{equation}
\label{eq:fseg!}
  f_{\rm pref}(t) = \left\{
  \begin{array}{lll}
  0.0 & {\rm for} & t < t_{\rm pref} , \\
  0.4 & {\rm for} & t \geq t_{\rm pref} , \\
  \end{array}\right.
\end{equation}
with $t_{\rm pref}= 0.2 \tdis$. A more gradual evolution would follow the simulations somewhat better, but a step function serves as a good approximation {(see \citet{lamers06} and Sect.~\ref{sec:disc})}.

{The assumption of constant $t_{\rm pref}/\tdis=0.2$ serves as a typical timescale after which the preferential loss of low-mass stars can become important. It can be justified by considering the mass-dependences of $t_{\rm pref}$ and $\tdis$. The total disruption time scales with cluster mass similarly to the dissolution timescale, i.e., $\tdis\propto M^{0.62}$. On the other hand, $t_{\rm pref}$ can be expected to scale with the half-mass relaxation time $t_{\rm rh}$, i.e., $t_{\rm pref}\propto M^{0.5}r_{\rm h}^{1.5}$. If we adopt a mean mass-radius relation $r_{\rm h}\propto M^{0.10\pm 0.03}$ \citep{larsen04b}, this leads to $t_{\rm pref}\propto M^{0.65}$, which is comparable to the mass-dependence of the total disruption time. We also considered models with $t_{\rm pref}=t_{\rm rh}$, assuming the same mass-radius relation, which indeed yields values for $t_{\rm pref}$ that are similar to $0.2\tdis$. 

Furthermore, we do not consider primordial mass segregation ($t_{\rm pref}=0$) for the model runs that are presented in this section. However, from other model runs where we did set $t_{\rm pref}=0$ we know that the effects of the preferential loss of low-mass stars on cluster observables are about 10\% stronger for primordial mass segregation than for the described results. Of course, $t_{\rm pref}$ can always be adapted to describe different forms of mass segregation.

The formulation in Eq.~\ref{eq:fseg!} implies that after $t_{\rm pref}$ dynamical mass loss simultaneously occurs in both modes.} Hereafter, any reference to the `{preferential loss of low-mass stars}' implies the use of this prescription for $f_{\rm pref}$ in our models. The value of $f_{\rm pref}(t\geq t_{\rm pref})$ is different from the one presented by \citet{lamers06} because they did not compare its value to $N$-body simulations: only $t_{\rm pref}$ was treated in that study.

\begin{figure*}[t]
\resizebox{\hsize}{!}{\includegraphics{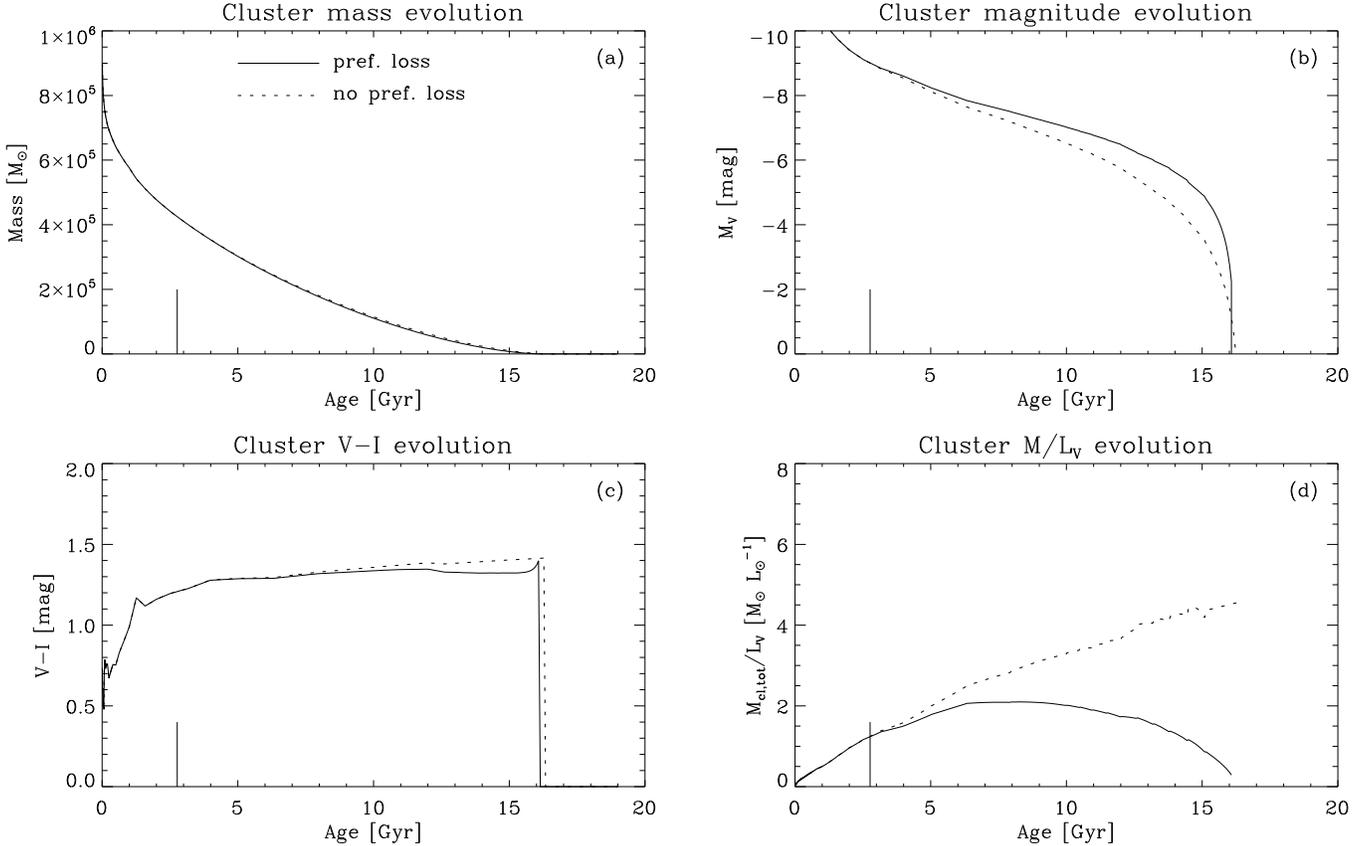}}
\caption[]{\label{fig:fxseg}
      Effect of {the preferential loss of low-mass stars} on (a) total cluster mass, (b) $V$-band magnitude, (c) $V-I$ and (d) $M/L_V$ evolution for clusters with initial mass $\mcli=10^6$~\msun, no stellar remnants, a dissolution timescale $t_0=3$~Myr {($\tdis=16$---16.5~Gyr)}, metallicity $Z=0.02$, and a Kroupa IMF. Solid curves denote cluster evolution when {the preferential loss of low-mass stars is included}, while clusters without {the preferential loss of low-mass stars} are represented by dotted lines. The onset of {the preferential mode $t_{\rm pref}$} is denoted by a vertical line.
    }
\end{figure*}
In Fig.~\ref{fig:fxseg} the effect of {the preferential loss of low-mass stars} on the total cluster mass $\mctot$, $V$-band magnitude, $V-I$ and mass-to-light ratio ($M/L_V$) evolution is shown for clusters with initial mass $\mcli=10^6$~\msun, no inclusion of stellar remnants (i.e., all remnants are immediately removed), a dissolution timescale $t_0=3$~Myr, metallicity $Z=0.02$, and a Kroupa IMF. This leads to a total disruption time of $\tdis\approx 16$~Gyr. For cluster evolution without {the preferential loss of low-mass stars}, some expected trends are immediately evident: mass and magnitude decrease with time, while $M/L_V$ increases and the cluster becomes redder.

\begin{figure*}[t]
\resizebox{\hsize}{!}{\includegraphics{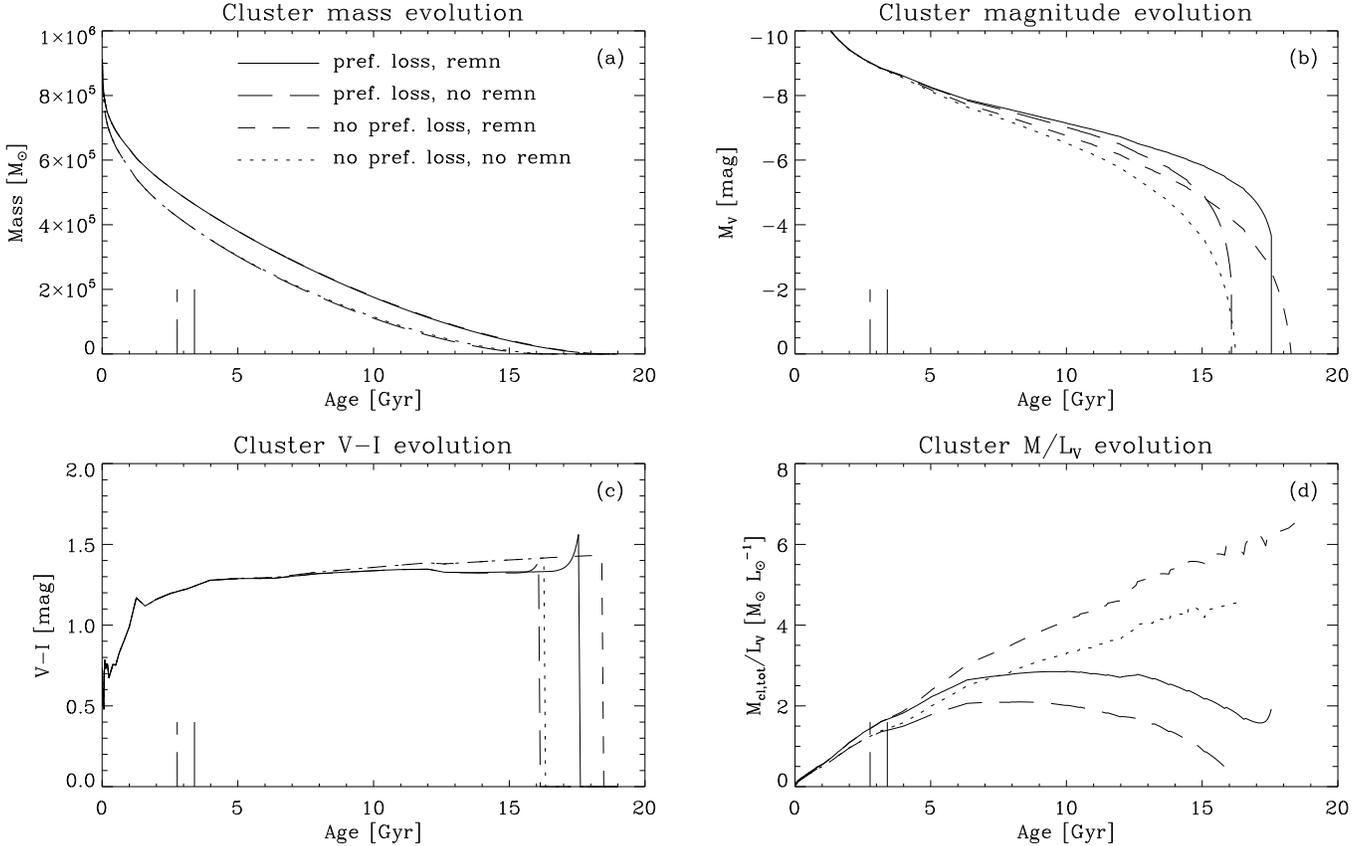}}
\caption[]{\label{fig:fxsr}
      Effect of stellar remnants on (a) total cluster mass, (b) $V$-band magnitude, (c) $V-I$ and (d) $M/L_V$ evolution for clusters with initial mass $\mcli=10^6$~\msun, a dissolution timescale $t_0=3$~Myr {($\tdis=16$---18.5~Gyr)}, metallicity $Z=0.02$, and a Kroupa IMF. Solid curves denote the evolution for clusters {including the preferential loss of low-mass stars} with stellar remnants, long-dashed curves without stellar remnants. For clusters without {the preferential loss of low-mass stars}, short-dashed curves represent the case in which remnants are included and dotted curves represent the result without stellar remnants. The onset of {the preferential mode $t_{\rm pref}$} is marked by vertical lines in the linestyles of the corresponding model runs.
    }
\end{figure*}

As can be observed when comparing the curves in Fig.~\ref{fig:fxseg}, {including the preferential loss of low-mass stars} has several implications for the cluster evolution {computed with our models}.
\begin{itemize}
\item[(1)]
{\it The total disruption time of a cluster {including the preferential loss of low-mass stars} hardly changes but is slightly smaller than for a cluster for which it is omitted}. {This is the result of our model assumptions, as the preferential loss of low-mass bodies causes a larger number of massive stars to be retained in the cluster.} The corresponding shorter lifetimes induce an increase in the cluster mass loss by stellar evolution. The enhanced decrease of total cluster mass consequently leads to a smaller total disruption time. {However, this effect does not include the possible decrease of the total disruption time due to quicker two-body relaxation in clusters with enhanced mean stellar masses.} Note that the decrease of total disruption time is of order $\sim 1$\%, which is best visible in the panel displaying colour evolution.
\item[(2)]
As soon as {the preferential loss of low-mass stars} starts, i.e., $f_{\rm pref}(t)>0$, {\it the cluster stays much brighter than for mass loss in the canonical mode}. Because the luminosity per unit mass is much higher for massive stars, a cluster {that preferentially loses low-mass stars} will be more luminous than a cluster of the same mass that does not. The change in the $V$-band magnitude induced by {the preferential loss of low-mass stars} peaks at about $0.9\tdis$ and is at most 1.5~mag.
\item[(3)]
The colour evolution of clusters is affected by {the preferential loss of low-mass stars} in two ways. {\it After the onset of energy equipartition, Fig.~\ref{fig:fxseg}(c) shows that these clusters are bluer than clusters without {the preferential loss of low-mass stars}, while just before total disruption reddening can be observed}. At first, the bottom end of the main sequence, which is being lost due to the preferential loss of low-mass stars, is redder than the average colour of the cluster. Due to the removal of its red constituents, such a cluster will appear bluer than a cluster without {the preferential loss of low-mass stars}. However, as stars are lost and $\mmin$ moves up the main sequence the colours of the stars that are ejected become increasingly blue. Cluster colour is then dominated by red giants and the ejected stars are bluer than the average cluster colour. Before total disruption, this induces a reddening of clusters {exhibiting the preferential loss of low-mass stars}. The ages at which these changes occur depend on cluster $\tdis$. For model runs with different $\tdis$ (either by varying initial cluster mass $\mcli$ or dissolution timescale $t_0$), we find that clusters {including the preferential loss of low-mass stars} always become slightly bluer from about $0.4\tdis$ on. However, the reddening is stronger and occurs at a smaller fraction of $\tdis$ for clusters with smaller total disruption times. This tendency is caused by massive red giants, which are redder and more luminous than low-mass giants. {For smaller total disruption times, the reddening causes clusters in the preferential mode to become much redder than clusters losing mass in the canonical mode during the last few percent of their lifetime. This effect is also present for the initial conditions used here if stellar remnants are included (see Fig.~\ref{fig:fxsr})}.
\item[(4)]
As can be observed in Fig.~\ref{fig:fxseg}(d), {\it the preferential loss of low-mass stars leads to a much smaller mass-to-light ratio than for clusters losing mass in the canonical mode}. Considering points (1) and (2), this is not suprising. A higher luminosity and slightly smaller mass together imply a decrease in $M/L_V$. The magnitude of the induced decrease is comparable to the change in cluster $V$-band luminosity, which follows from the magnitude change to be about a factor six.
\end{itemize}
In their study, \citet{lamers06} obtained magnitude evolution curves that are much more weakly affected by {the preferential loss of low-mass stars} than is shown in Fig.~\ref{fig:fxseg}. In the present paper, cluster photometries are directly computed from the changing stellar mass function, which is a more direct method than the one used in \citet{lamers06}\footnote{\citet{lamers06} calculated magnitudes by using the photometry of stellar mass-truncated {\it GALEV} models.}. Therefore, our computation of cluster evolution provides an update to their results. Furthermore, they found that clusters {including the preferential loss of low-mass stars} are bluer for $0.4<t/\tdis<0.8$ and redder for $t\geq 0.8\tdis$. Our extended study of the parameter space shows that especially the value of $0.8$ depends on the total disruption time of the cluster.

The effect of {the preferential loss of low-mass stars} is clearly visible in {\it magnitude, colour} and {\it $M/L_V$ evolution} of clusters. Especially cluster magnitude and $M/L_V$ are significantly affected. This implies that when considering either observable, it is very important to detemine whether the cluster {exhibits the preferential loss of low-mass stars} or not, {for example by checking whether it is mass-segregated}. In order to obtain an appropriate interpretation of cluster colour, it has to be determined how close to total disruption the cluster is.

\subsubsection{The effects of stellar remnants} \label{sec:fxsr}
The inclusion of stellar remnants follows the description presented in Sects.~\ref{sec:remnprod} and~\ref{sec:diss}. {In our models, it implies that part of the mass is retained upon the death of a star and clusters thus lose mass due to stellar evolution at a slower rate.} It can be expected to affect cluster mass evolution, because a significant fraction of cluster mass can be constituted by remnants. Clearly, the mass-to-light ratio will then be altered as well. To assess the effect of stellar remnants on cluster evolution, our model results are shown in Fig.~\ref{fig:fxsr} for clusters with initial mass $\mcli=10^6$~\msun, with and without {the preferential loss of low-mass stars}, a dissolution timescale $t_0=3$~Myr {($\tdis=16$---18.5~Gyr)}, metallicity $Z=0.02$ and a Kroupa IMF.

We observe the following changes {in our models} due to the inclusion of stellar remnants.
\begin{itemize}
\item[(1)]
Because the net mass loss due to stellar evolution is smaller if stellar remnants are retained in the cluster after the death of their progenitors, {\it the total cluster mass is higher than in the non-remnant case at any time}. The immediate consequence is {that the models predict} a larger total disruption time. For clusters {including the preferential mode}, which more easily keep their remnants (see Sect.~\ref{sec:dissseg} and point (3)), this effect is smaller than for clusters without {the preferential loss of low-mass stars}, because the death criterion of clusters in our simulations ($\mcl<100~\msun$) only depends on luminous cluster mass. Retaining remnants implies that mass loss due to dissolution more strongly affects luminous cluster mass, causing a cluster that exhibits {the preferential loss of low-mass stars} to satisfy the death criterion {of our models} earlier than expected for its total mass. However, though this weakens the lifetime-increasing effect of keeping stellar remnants, it never dominates.
\item[(2)]
For clusters without {the preferential loss of low-mass stars}, where the fraction of mass lost in the form of remnants is simply equal to the remnant mass fraction, the inclusion of remnants leads to a higher luminosity. Increasing luminosity by including remnants might be counter-intuitive. However, when including stellar remnants part of the cluster mass loss by dissolution is in the form of remnants instead of luminous stars. This implies that the average luminosity of bodies that are lost by dissolution is smaller than in the case without remnants, leading to a smaller luminosity decrease. Though less explicitly, the same effect is present in clusters {including the preferential loss of low-mass stars} since low-mass stars hardly contribute to the total cluster luminosity. Moreover, the lifetime-extending effect of remnants also implies that (luminous) mass is lost at a slower pace. We conclude that {\it the cluster luminosity decrease due to dissolution becomes smaller if stellar remnants are included}.
\begin{figure*}[t]
\resizebox{\hsize}{!}{\includegraphics{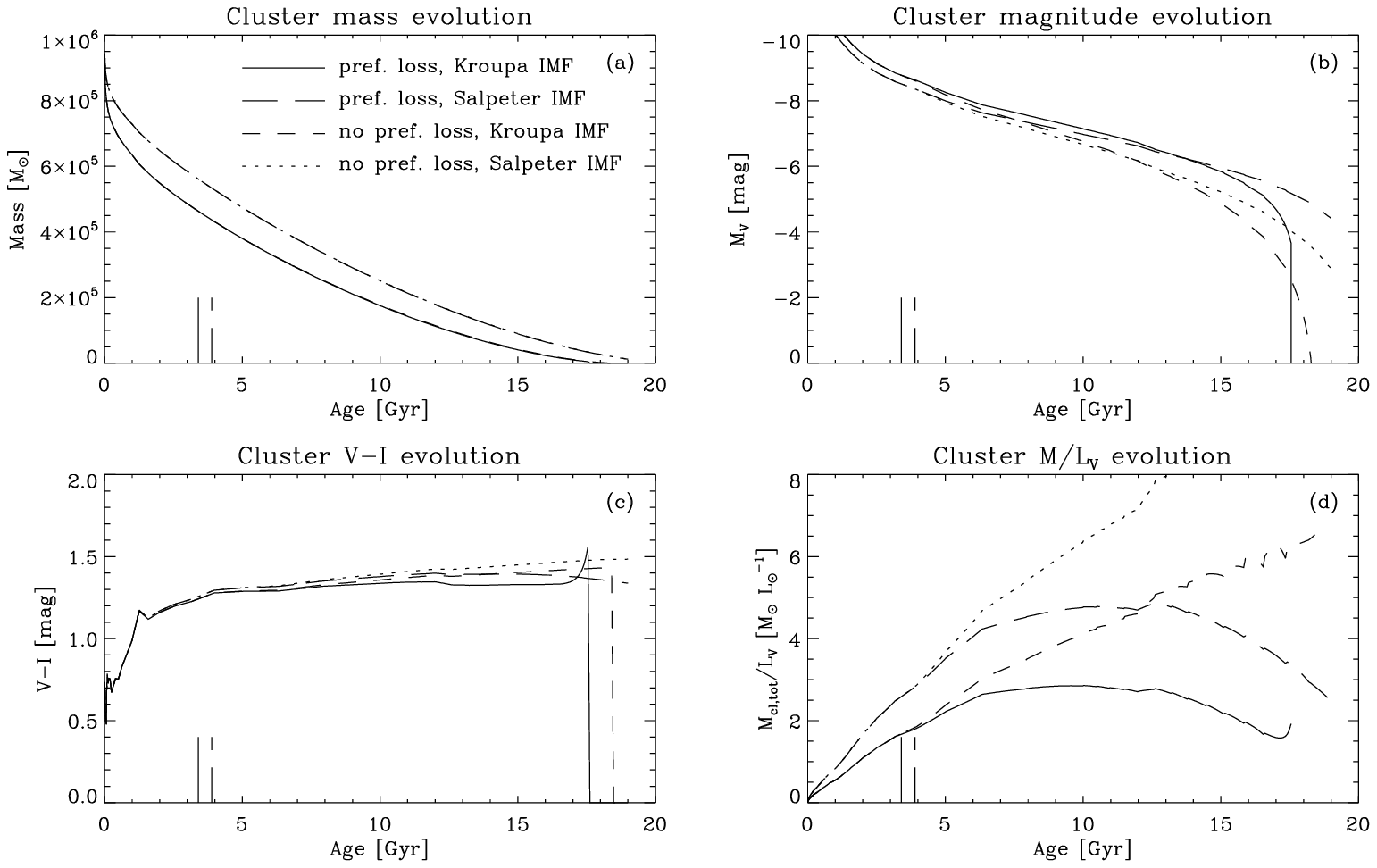}}
\caption[]{\label{fig:fximf}
      Effect of initial mass function on (a) total cluster mass, (b) $V$-band magnitude, (c) $V-I$ and (d) $M/L_V$ evolution for clusters with initial mass $\mcli=10^6$~\msun, including stellar remnants, a dissolution timescale $t_0=3$~Myr {($\tdis=17$---21~Gyr)} and metallicity $Z=0.02$. For a Kroupa IMF, solid curves denote the evolution for clusters {including the preferential loss of low-mass stars}, short-dashed lines for clusters with canonical mass loss. Clusters with a Salpeter IMF are represented by long-dashed lines when {the preferential loss of low-mass stars} is included and dotted lines describe the case where it is omitted. The onset of {the preferential mode $t_{\rm pref}$} is marked by vertical lines in the linestyles of the corresponding model runs.
    }
\end{figure*}
\item[(3)]
The mass-to-light ratio evolution shows a very clear effect of stellar remnants as {\it for both mass loss modes the $M/L_V$ curves are much higher if remnants are included}. This means that the relative increase of cluster mass due to remnants is larger than the corresponding relative increase of cluster magnitude that was discussed at point (2). The former is a direct consequence of adding remnants, while the latter is an induced effect: the average $M/L_V$ of all bodies in a cluster is higher per definition if dark mass is added. Furthermore, {\it for clusters with the preferential loss of low-mass stars that also includes stellar remnants, the mass-to-light ratio shows an increase during the final part of cluster evolution}. This can be attributed to the preservation of remnants in clusters losing mass {in the preferential mode}, which is due to the fact that remnants can only be lost from these clusters if $\mssr<\mmin$.
\end{itemize}
Because remnant production or loss does not directly alter the colour composition of luminous cluster content, colour evolution is hardly affected by the inclusion of remnants.

We find that the inclusion of stellar remnants strongly affects the {\it mass, magnitude} and {\it $M/L_V$ evolution} of clusters. This is because part of the cluster mass loss occurs in the form of remnants rather than luminous stars. The extent of the differences (up to 30\% at $t=0.5\tdis$ and increasing afterwards) suggests that a proper treatment of remnants should be included in any cluster model.

\subsection{The effects of the stellar IMF} \label{sec:fximf}
As discussed in Sect.~\ref{sec:diss}, our models can be calculated for any multi-component power law IMF. {Different IMFs are likely to exhibit a tendency to higher or lower stellar masses with respect to one another.} For the Kroupa and Salpeter IMFs, the consequences of this effect are investigated here. As mentioned in Sect.~\ref{sec:clevo}, the lower mass limit is taken to be $\mmini=0.08~\msun$ for a Kroupa IMF {and $\mmini=0.1~\msun$ for a Salpeter IMF}. Since a Salpeter IMF has a slightly steeper slope than a Kroupa IMF, and the latter features a bend at $0.5$~\msun, the Salpeter IMF has a lower mean stellar mass. Figure~\ref{fig:fximf} displays our results for both IMFs in the case of clusters with initial mass $\mcli=10^6$~\msun, with and without {the preferential loss of low-mass stars}, with stellar remnants, a dissolution timescale $t_0=3$~Myr {($\tdis=17.5$---21~Gyr)} and metallicity $Z=0.02$.

The initial mass function affects the resulting cluster evolution in the following ways.
\begin{itemize}
\item[(1)]
{\it Total cluster mass stays higher for clusters with a Salpeter IMF}. Consequently, their total disruption times are increased as well. Since it has a slightly steeper slope, the mass loss due to stellar evolution is smaller than for a Kroupa IMF. The resulting higher cluster mass leads to dissolution acting on a longer timescale, thereby also contributing to an extended lifetime of the cluster. {In this case, the example model clusters with a Salpeter IMF survive beyond the maximum age spanned by the models.}
\begin{figure*}[t]
\resizebox{\hsize}{!}{\includegraphics{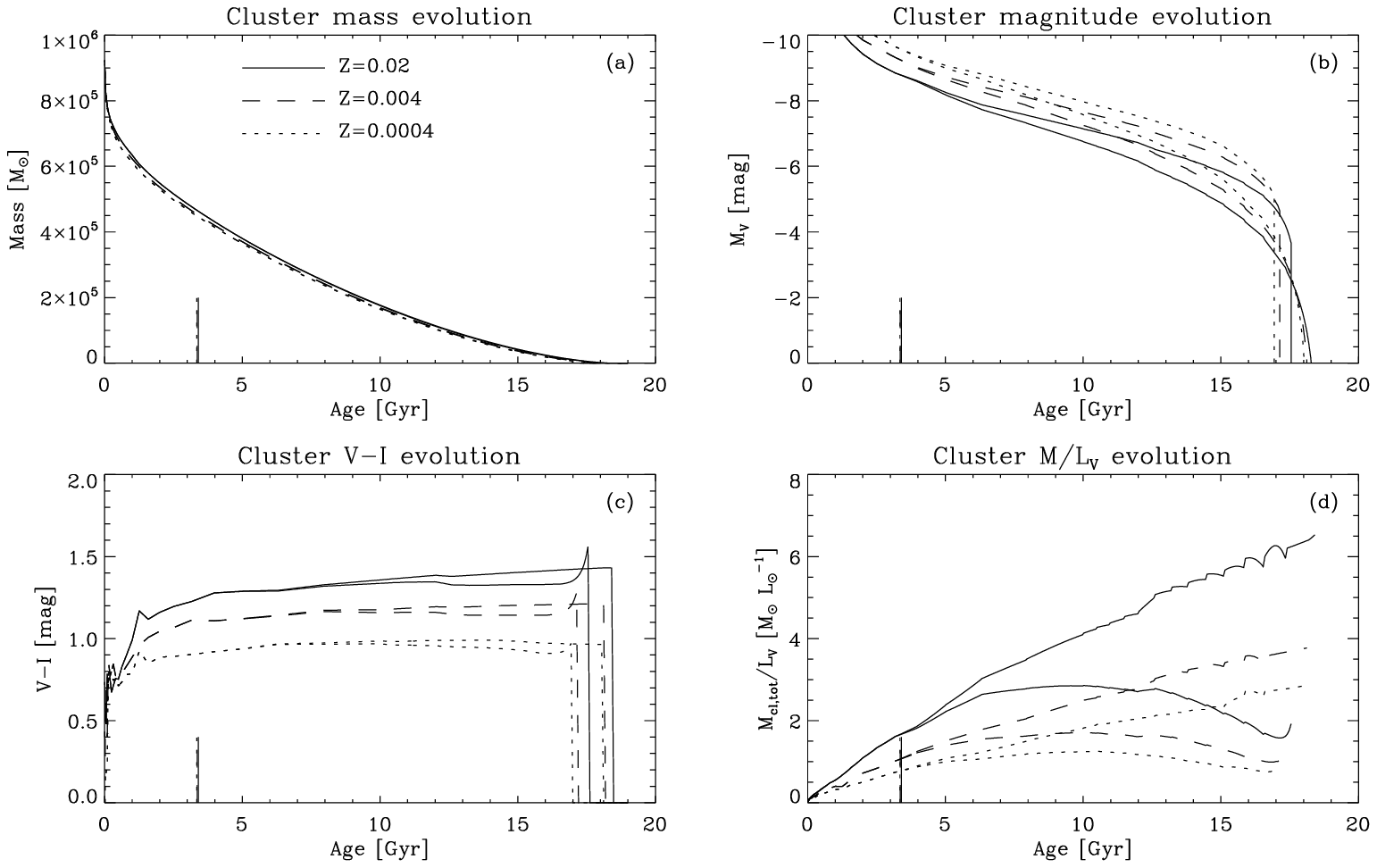}}
\caption[]{\label{fig:fxz}
      Effect of metallicity on (a) total cluster mass, (b) $V$-band magnitude, (c) $V-I$ and (d) $M/L_V$ evolution for clusters with initial mass $\mcli=10^6$~\msun, including stellar remnants, a dissolution timescale $t_0=3$~Myr {($\tdis=17$---18.5~Gyr)} and a Kroupa IMF. Solid curves denote the evolution for $Z=0.02$, dashed curves for $Z=0.004$, and dotted ones for $Z=0.0004$. Results with and without {the preferential loss of low-mass stars} show effects as presented in Fig.~\ref{fig:fxseg} and described in Sect.~\ref{sec:fxseg} and are therefore represented by the same linestyles. The onset of {the preferential mode $t_{\rm pref}$} is marked by vertical lines in the linestyles of the corresponding model runs.
    }
\end{figure*}
\item[(2)]
Because it favours stars of higher masses, {\it a Kroupa IMF leads to clusters that are initially slightly brighter than for a Salpeter IMF}. However, the mass decrease due to stellar evolution is also stronger for a Kroupa IMF, causing its cluster mass to be lower than for a Salpeter IMF. As a result, {\it the luminosities of clusters with a Salpeter IMF overtake those with a Kroupa IMF later on}. Though still very weak, the transition is best visible when comparing the curves {corresponding to clusters with mass loss in the preferential mode} in Fig.~\ref{fig:fximf}(b). For these, the transition occurs at about $0.7\tdis$. From model runs with other total disruption times, we observe that this fraction of $\tdis$ is not constant. It significantly increases for smaller $\tdis$, with $M_V$ of a Salpeter IMF always being fainter than that of a Kroupa IMF for $\tdis\simless 500$~Myr.
\item[(3)]
{\it Clusters with a Kroupa IMF are slightly bluer than those with a Salpeter IMF}. However, this effect is too small to be observable in real clusters. It can be understood by considering the relatively larger contribution of massive (blue) stars in evolved clusters with a Kroupa IMF. {Had the Salpeter examples been disrupted within the model age range, the characteristic reddening just before total disruption would have been visible for that IMF as well.}
\item[(4)]
The higher masses of clusters with a Salpeter IMF and their {generally lower} luminosities that were described at point (2) together lead to {\it higher mass-to-light ratios for clusters evolving from a Salpeter IMF}. {Again, for clusters {including the preferential loss of low-mass stars}, the final $M/L_V$ increase upon total disruption would also be visible for the Salpeter IMF if it would have been completely disrupted within the model age range}.
\end{itemize}
Although specifically applied to the Kroupa and Salpeter IMFs, qualitatively the results of the above analysis hold for any two mass functions of which one has a different mean mass than the other. Quantitatively, there will still be variations depending on the specific IMF.

Because the underlying stellar IMF in a cluster determines its future mass loss due to stellar evolution, it strongly affects the {\it total disruption time} and {\it mass-to-light ratio}. Any treatment of these two observables requires an accurate description of the IMF.

\subsection{The effects of metallicity} \label{sec:fxz}
To investigate the influence of metallicity on our results, we consider $Z=\{0.0004, 0.004, 0.02\}$ in Fig.~\ref{fig:fxz}. Model results are shown for clusters with initial mass $\mcli=10^6$~\msun, with and without {the preferential loss of low-mass stars}, including stellar remnants, a dissolution timescale $t_0=3$~Myr {($\tdis=17$---18.5~Gyr)} and a Kroupa IMF.

The effects of metallicity on the results are as follows.
\begin{itemize}
\item[(1)]
We see that {\it total cluster mass is hardly affected by metallicity at any time}. It marginally increases with $Z$, which is caused by more rapid stellar evolution for low metallicities \citep[e.g.,][]{hurley00,hurley04}. Consequently, the total disruption time also slightly increases with metallicity, which is best observed in the $M_V$ and $V-I$ panels. Nevertheless, the effect is small, less than 10\% of the total lifetime. This is in excellent agreement with the results from \citet{hurley04}.
\item[(2)]
{\it Low-$Z$ clusters are brighter than high-$Z$ clusters}. This is due to a general luminosity decrease of stars with metallicity \citep[e.g.,][]{girardi00,hurley04}. The difference is observed for clusters with and without {the preferential loss of low-mass stars}, and is typically more than one $V$-band magnitude.
\item[(3)]
The colour evolution shows a uniform trend with $Z$. {\it Clusters with high metallicity are much redder than clusters with low metallicity}. This is the result of stellar atmospheres and stellar evolution \citep[e.g.,][]{hurley00}. For high ages, the $V-I$ value is more or less constant for each metallicity, underlining its value as metallicity probe for globular clusters when considering broadband colours \citep[e.g.,][]{maraston05}. The $V-I$ shift between $Z=0.0004$ and $Z=0.02$ is about 0.5~mag, but varies for colours at other wavelengths. This is in accordance with the fact that clusters of different metallicities move on clearly distinguishable paths in colour-colour diagrams \citep[e.g.,][]{bruzual03}.
\item[(4)]
The higher luminosity of low-metallicity stars and the consequently slightly enhanced mass loss by stellar evolution induce a common effect on cluster mass-to-light ratios. From Fig.~\ref{fig:fxz}(d) we see that {\it $M/L_V$ strongly increases with metallicity}. The effect is strong enough to move the $M/L_V$ evolution of a $Z=0.0004$ cluster {with canonical mass loss} through the $M/L_V$ range of a cluster that does {include the preferential loss of low-mass stars} at $Z=0.02$. This apparent degeneracy is lifted by taking cluster colours into account (see point (3)).
\end{itemize}
Whenever cluster {\it magnitude, colour} and {\it mass-to-light ratio} evolution are considered, cluster metallicity plays an important role. They are all strongly affected by the adopted metallicity. For globular cluster populations this implies that a range of colours and mass-to-light ratios can be covered by a metallicity spread of the population or internal cluster processes like self-enrichment.

In \citet{lamers05}, an expression is provided for the total disruption time $\tdis$ as a function of $t_0$, $\mci$ and $\gamma$. In this study, we find that $\tdis$ depends on the inclusion of {the preferential loss of low-mass stars}, stellar remnants, IMF and metallicity. Therefore, values for $\tdis$ are best obtained by integrating the models presented in this paper. Regardless, the expression from \citet{lamers05} can still be used to estimate $\tdis$ with approximately 20\% accuracy.

\section{Application to globular clusters} \label{sec:appl}
The results presented in Sect.~\ref{sec:results} cover a wide range of masses, magnitudes, colours and mass-to-light ratios. As a first indication, it is relevant to check whether the properties of Galactic globular clusters can be reproduced in our models. For this purpose, the results have to be considered at ages $t\approx 12$~Gyr. From \citet{harris96} the $M_V$ range is found to be $M_V=-1.60$ (Pal~1) to $M_V=-10.29$ ($\omega$Cen). For the Solar neighbourhood value of the dissolution timescale in the case of $e=0.7$ orbits $t_0=3$~Myr\footnote{See Sect.~\ref{sec:fxseg}.}, we find that this range can be covered at $t=12$~Gyr for any metallicity $Z\leq 0.02$ if the maximum {\it initial} cluster mass equals $M_{\rm cl}^{\rm max}=10^7$~\msun. For longer dissolution timescales (i.e., larger galactocentric radii) the observed magnitude range can be covered with even smaller maximum initial cluster masses. {Please note that the dissolution timescale depends on the tidal field strength and that it therefore varies for different globular cluster orbits. This implies that it is not possible to impose limits on the properties of the complete globular cluster population (like their maximum initial masses) from an analysis in which this variation is not incorporated.}

\subsection{The mass-to-light ratio}
Galactic globular cluster mass-to-light ratios are found to be $M/L_V=1.45\pm 0.1~\msun~{\rm L}_\odot^{-1}$ \citep{mclaughlin00}. SSP models, in which only stellar evolution is included and dynamical effects are neglected, predict a {\it minimum} value of $M/L_V\approx 2~\msun~{\rm L}_\odot^{-1}$ at $t=12$~Gyr, requiring the minimum metallicity of our models $Z=0.0004$ (see Fig.~\ref{fig:fxz}). This further complicates explaining the observed mass-to-light ratios, since globular cluster metallicities are typically $Z=0.0004$---$0.014$ \citep{vandalfsen04}. Dynamical effects are thus needed to explain the even smaller mass-to-light ratio of Galactic globular clusters. If {the preferential loss of low-mass stars} is included, cluster mass-to-light ratio curves do span the correct part of $M/L_V$ space, ranging down to $M/L_V<1~\msun~{\rm L}_\odot^{-1}$  \citep[see also][]{kruijssen08b}. From Fig.~\ref{fig:fxsr} we see that accounting for stellar remnants yields an {\it increase} up to $1~\msun~{\rm L}_\odot^{-1}$ relative to model clusters without stellar remnants, obviously implying that they should be included for accurate interpretations of globular cluster observations. If globular clusters are populated using a Salpeter IMF rather than a Kroupa IMF, this effect is nearly doubled (see Fig.~\ref{fig:fximf}). From Fig.~\ref{fig:fxz}(d) we find that cluster models at high metallicities do not reproduce the low observed $M/L_V=1.45\pm 0.1~\msun~{\rm L}_\odot^{-1}$. Thus, metal-poor clusters {including the preferential loss of low-mass stars} with a Kroupa IMF\footnote{Or any other IMF that slightly favours massive stars with respect to a Salpeter IMF.} are required to reach the low average mass-to-light ratio. If these conditions are met, the moderately flat $M/L_V$ evolution curves show that a more or less constant time-average is not surprising. 

Some globular clusters have mass-to-light ratios that are much higher than the mean value of $M/L_V=1.45\pm 0.1~\msun~{\rm L}_\odot^{-1}$. An example is $\omega$Cen \citep[$M/L_V=2.5~\msun~{\rm L}_\odot^{-1}$, see][]{vandeven06}, which is also the most massive Galactic globular cluster\footnote{In fact, $\omega$Cen is not a normal globular cluster since {there are strong indications that it could be} a stripped dwarf galaxy \citep[e.g.,][]{ideta04}.}, with $M_{\rm cl}^{\rm \omega Cen}=2.5\times 10^6$~\msun \citep{vandeven06}. Since a high mass implies a large relaxation time, this could agrees with the view that $\omega$Cen has not yet {reached energy equipartition}. 
\begin{figure*}[t]
\center
\resizebox{16cm}{!}{\includegraphics{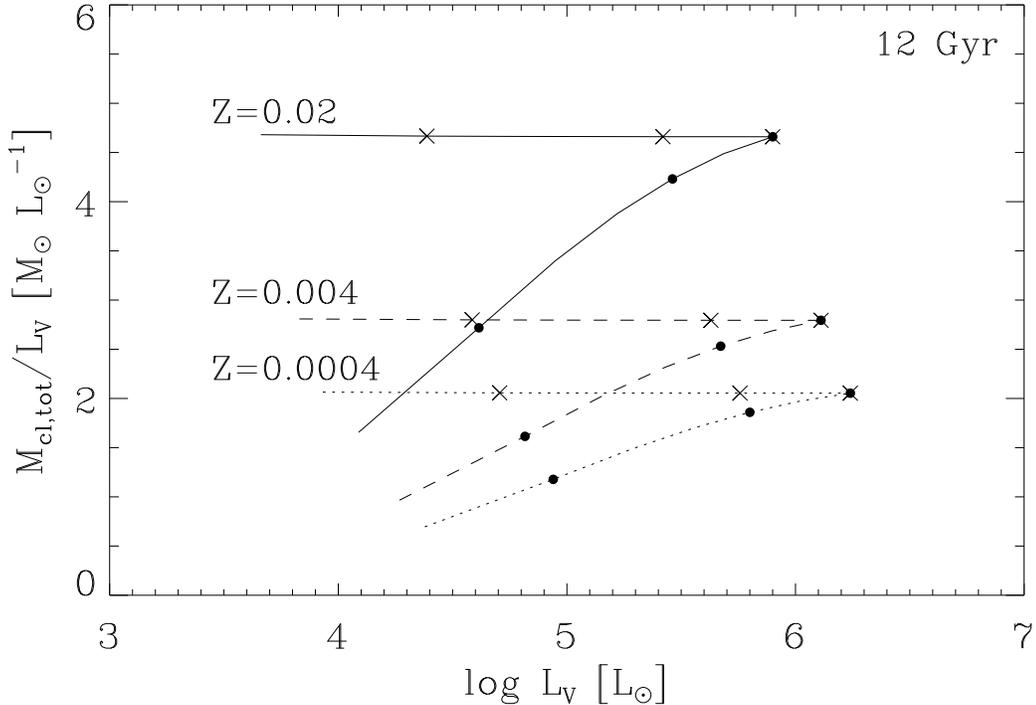}}
\caption[]{\label{fig:mlcst}
      Effect of {the preferential loss of low-mass stars} on the relation between mass-to-light ratio and luminosity at an age of $t=12$~Gyr. Metallicities $Z=\{0.0004, 0.004, 0.02\}$ are denoted by dotted, dashed and solid curves, respectively. Horizontal lines represent clusters without {the preferential loss of low-mass stars}, while inclined curves show the relation if it is included. The models are computed for initial masses between 10$^2~\msun$ and 10$^7$~\msun, a Kroupa IMF, dissolution timescale $t_0=3$~Myr and including stellar remnants. On each curve, clusters with (from left to right) {\it initial} masses $\log{\mcli}=\{6,6.5,7\}$ are marked with crosses (canonical mode) and dots (preferential mode). The present day mass can be derived from the luminosity and mass-to-light ratio.
    }
\end{figure*}

Another important implication of our analysis of mass-to-light ratio evolution is that globular clusters of comparable ages can {\it not} be assumed to have constant $M/L_V$ for different cluster luminosities at fixed metallicity. For a given age, the mass-to-light ratio can strongly ($\sim 0.6$~dex) depend on the dynamical state of the cluster (see Fig.~\ref{fig:fxseg}(d)), and thus on cluster mass and luminosity. The variation of $M/L_V$ with luminosity when including {mass loss in the preferential mode} is illustrated in Fig.~\ref{fig:mlcst}. Models are shown of clusters with and without {the preferential loss of low-mass stars}, metallicities $Z=\{0.0004, 0.004, 0.02\}$ (dotted, dashed and solid curves, respectively), age $t=12$~Gyr, a maximum cluster mass of 10$^7$~\msun, a Kroupa IMF, dissolution timescale $t_0=3$~Myr and including stellar remnants. The horizontal lines denote the constant mass-to-light ratios predicted if {the preferential loss of low-mass stars} is ignored, while the inclined curves show the relation between $M/L_V$ and luminosity if it is accounted for.

Since more luminous clusters are also more massive, the onset of {mass loss in the preferential mode} occurs later on for these clusters, implying that its effects are weaker for massive clusters at any age. Because {the preferential loss of low-mass stars} decreases the mass-to-light ratio, this decrease is thus smaller for clusters of higher masses, leading to a mass-to-light ratio that increases with cluster mass and $L_V$ as in Fig.~\ref{fig:mlcst}. Observational evidence of this effect for the same quantitative range has been found for Galactic and extragalactic globular clusters \citep[e.g.,][]{mandushev91,rejkuba07,kruijssen08b}. If cluster masses are determined using a fixed $M/L_V$, thereby not accounting for the effects of {the preferential loss of low-mass stars}, these masses can be strongly {\it overestimated} by as much as 0.6~dex. Because the error is larger for lower masses, the slope of the inferred cluster mass function will be {\it underestimated} (i.e., a negative slope will be steeper) if {the preferential loss of low-mass stars} is ignored. 

{It is straightforward to derive a quantitative estimate for the effect on the inferred (powerlaw) cluster mass function. Let us consider clusters with `true' mass $M$ that are exhibiting the preferential loss of low-mass stars and are thus in the regime that is inconsistent with canonical cluster models. If we now use a powerlaw with index $A$ to reasonably approximate the mass-to-light ratio increase with luminosity from Fig.~\ref{fig:mlcst}, i.e., $M/L\propto L^A$ for $\log{L}\simless 5.5$ depending on metallicity, then the ratio of the photometrically inferred mass $M_{\rm SSP}$ for which constant $M_{\rm SSP}/L$ is assumed to its true mass $M$ scales as $M_{\rm SSP}/M\propto L^{-A}$. This is equivalent to using a powerlaw with index $B=A/(1+A)$ to approximate the mass-to-light ratio increase with {\it true mass}, i.e., $M/L\propto M^B$, leading to $M_{\rm SSP}/M\propto M^{-B}$. Then, for a `true' slope of the cluster mass function $-\alpha$, its {\it photometrically inferred} slope $-\alpha_{\rm SSP}$ using constant mass-to-light ratios is given by 
\begin{equation}
\label{eq:alpha}
-\alpha_{\rm SSP}=-\alpha-\alpha A+A=(-\alpha+B)/(1-B).
\end{equation}
From Fig.~\ref{fig:mlcst} we find that typically $A\sim0.27$ (and thus $B\sim 0.21$), implying that for $\alpha=2$ we find $\alpha_{\rm SSP}\sim 2.27$. The deviation increases with the age of the cluster sample (here we used $t=12$~Gyr, while for $t=3$~Gyr $\alpha_{\rm SSP}\sim 2.23$) and the above approach assumes a constant dissolution timescale for the cluster sample. For (more realistic) varying dissolution timescales, the deviation will typically be between 50\% and 100\% of the the presented value. This is still significant and should thus be accounted for when studying the cluster mass function of old cluster samples.}

The variation of $M/L_V$ with luminosity is crucial for cluster masses inferred from luminosities, which can be incorrectly determined by 0.6~dex. Because of increasing evidence that more globular clusters {are preferentially losing low-mass stars} and might have evolved to a mass-segregated state than previously thought \citep[e.g.,][]{demarchi07}, this effect should always be considered when studying globular clusters. 

\subsection{Colours of globular clusters}
The typical colour range of Galactic globular clusters is $0.75$---$1.1$ in $V-I$ \citep{smith07}. From Fig.~\ref{fig:fxz}(c) we can conclude that this range is covered by varying metallicity in our models. Obviously, IMF variations might slightly affect the colour range as well, with top-heavy IMF showing a tendency to bluer colours (see Fig.~\ref{fig:fximf}(c)). However, this effect is only significant for strongly deviating IMFs. Still, from all figures in Sect.~\ref{sec:results} we still see that colours are affected by {the preferential loss of low-mass stars} and the colour range is increased. Figure~\ref{fig:fxz}(c) illustrates that if {mass loss in the preferential mode} would be ignored, uncertainties up to 0.2~mag (0.1~mag up and down) should be included for computed cluster colours at a any fixed age and metallicity. This happens to be the same colour spread at fixed metallicity as found by \citet{smith07} in their colour-metallicity relation.

\subsection{Ultra-compact dwarf galaxies}
We see that including {the preferential loss of low-mass stars} and stellar remnants allows for a more extensive analysis of globular clusters and affects their reproduced property ranges. It is evident that globular clusters or more massive globular cluster-like objects with very high mass-to-light ratios (with regard to their metallicities) {cannot have experienced significant mass loss in the preferential mode} unless their $M/L_V$ is increased by a strongly differing IMF or by agents unaccounted for by our models, for instance by intermediate mass black holes, modified gravity or dark matter. Hence, if indications for {the preferential loss of low-mass stars} (such as mass segregation or a bottom-depleted mass function) are found for such objects, they points to these causes for the high observed $M/L_V$. {For objects with masses $M>10^7~\msun$ this is unlikely to occur, since their relaxation times are of the order of a Hubble time or larger.}

If there is a present day globular cluster mass above which {no clusters have yet reached energy equipartition}, any objects above that mass will have high mass-to-light ratios with respect to clusters below that mass. {This is found for UCDs, of which the $M/L_V$-range is typically 2---10~$\msun~{\rm L}_\odot^{-1}$ \citep[e.g.,][]{mieske08a}}. Pending the role of dark matter these galaxies could be regarded a natural continuation of the mass spectrum beyond globular clusters. Indications for such a continuation are found by \citet{wehner07} for UCD candidates in NGC~3311. {As is clear from Fig.~\ref{fig:fximf}(d), our models without the preferential loss of low-mass stars produce mass-to-light ratios similar to UCDs if a Salpeter IMF is assumed \citep[in agreement with][]{hilker07}, while the canonical Kroupa IMF cannot reproduce the range that is typical to UCDs \citep[as is also found by][]{mieske08}. Nonetheless, any possible connection between globular clusters and UCDs would be smoothened by metallicity spreads and the possible influence of dark matter}, making it a challenge to be directly observed.

{Contrary to low-mass UCDs, more massive examples have not reached energy equipartition within a Hubble time due to their long relaxation times. However, this would lead to constant mass-to-light ratios at high masses, while \citet{rejkuba07} have shown that the trend of increasing $M/L_V$ with mass is even stronger for UCDs than it is for globular clusters.}

\section{Discussion} \label{sec:disc}
We have described models to calculate the evolution of star clusters including {the preferential loss of low-mass stars} and stellar remnants. Furthermore, they have been used to investigate the influence of these model components, as well as of the IMF and metallicity on cluster evolution. In this section we discuss the assumptions that were made, their influence on the results and the applicability of our models. We also indicate how the models can be improved.

\subsection{Influence of assumptions} \label{sec:assump}
The models presented in this study are based on the following assumptions.
\begin{itemize}
\item[(1)]
We adopt the Padova 1999 stellar evolution models (Bertelli et al. 1994, AGB treatment as in Girardi et al. 2000) to compute photometry, determine stellar lifetimes and describe the consequent cluster mass loss due to stellar evolution. For the {\it mass evolution} of all stars we assume constant stellar masses until their instantaneous deaths. Only very massive stars experience strong mass loss during a significant part of their lives, but these stars hardly contribute to the total cluster mass. Low-mass stars with $\ms\simless 8$~\msun~only suffer significant mass loss during the last 10\% of their lives. Therefore, instantaneous death is a reasonable assumption when calculating cluster mass. {\it The photometric properties of stars are not affected} by our assumption of instantaneous death, since stellar photometry as described in the Padova models includes the photometric effects of stellar mass loss. The Padova isochrones include a description of AGB evolution, which is very important since AGB stars dominate the photometric evolution of stars after about 100~Myr.
\item[(2)]
If stellar remnants are included, upon its death a star is replaced by a body with mass determined by the initial-remnant mass relation. For white dwarfs, the relation from \citet{kalirai07} is assumed, while neutron star and black hole masses are based on studies by \citet{nomoto88} and \citet{casares06}, respectively. Our resulting initial-remnant mass relation for the full stellar mass range is independent of metallicity. However, generally a metallicity dependence is found for both white dwarf \citep[e.g.,][]{kalirai05,meng07} and neutron star masses \citep[e.g.,][]{hurley00}. Remnants formed at high metallicities are generally less massive than those formed at low metallicities. The difference between $Z=0.0004$ and $Z=0.02$ is typically $\sim 10$\%, which implies that the speed of the increase of non-luminous cluster mass $\msr$ due to stellar remnant production would be enhanced by 10\% at low metallicities. For Fig.~\ref{fig:fxz}, which shows the effect of metallicity on our results, this would have some small consequences. Assuming a remnant to total cluster mass fraction $\msr/\mctot\approx 0.5$, the slope of total cluster mass evolution (Fig.~\ref{fig:fxz}(a)) would be increased (i.e., become less steep) by a few percent at low metallicities, extending the total disruption time by a comparable but slightly lower percentage due to the $\mctot^\gamma$ mass dependence of the dissolution timescale. Moreover, curves describing $M/L_V$ cluster evolution (Fig.~\ref{fig:fxz}(d)) would exhibit metallicity effects that are smaller by a few percent. Overall, these corrections are not sufficiently large to have a significant effect on our results.
\item[(3)]
We assume all remnant bodies to be initially bound to the cluster. In reality, supernovae may induce kick velocities on their remnant {black holes and} neutron stars \citep[e.g.,][]{portegieszwart97}. A fraction of these remnants will have velocities larger than the escape velocity. Typical escape velocities of globular clusters are $\sim 30$~km~s$^{-1}$ \citep{mclaughlin05}, while in some cases kick velocities of several hundreds km~s$^{-1}$ are observed \citep[e.g.,][]{lyne94}. This would imply that it would not be possible to retain all black holes and neutron stars in globular clusters, but nonetheless high numbers of neutron stars are observed in real globular clusters \citep{camilo00}. \citet{pfahl02} suggest that low kick velocities are obtained if neutron stars are formed in long-period and low-eccentricity high-mass X-ray binary (HMXB) systems. In that case, the retained neutron star fraction would be four times higher than expected for commonly observed `fast' neutron stars. Because of the lack of any definitive answer to the black hole and neutron star retention problem, and for the sake of model simplicity, we ignore kick velocities. In Fig.~\ref{fig:fxns} we show the effect of including or excluding all black holes and neutron stars from our models. The panel is identical to that of Fig.~\ref{fig:fxsr}, however curves representing model runs without stellar remnants are now replaced by ones which include white dwarfs {\it only} and for which all black holes and neutron stars have been removed. The effect is negligibly small (e.g., $\simless 10\%$ in all observables) because the total remnant mass is generally dominated by white dwarfs at ages $\simgreat 100$~Myr. In reality, the effect will be even smaller since a number of black holes and neutron stars have velocities smaller than the escape velocity and are thus retained in the cluster.
\begin{figure*}[t]
\resizebox{\hsize}{!}{\includegraphics{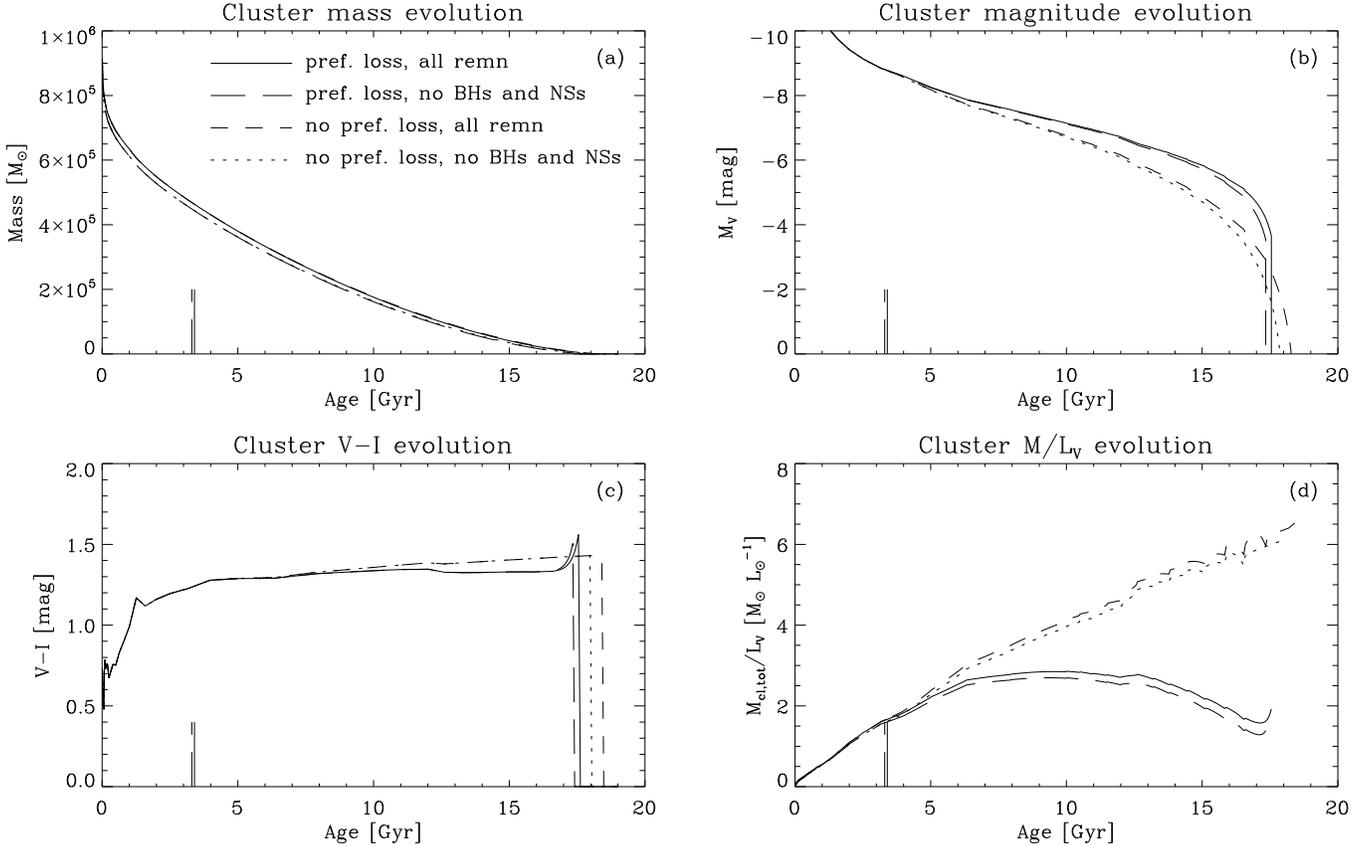}}
\caption[]{\label{fig:fxns}
      Effect of neutron star ejection on (a) total cluster mass, (b) $V$-band magnitude, (c) $V-I$ and (d) $M/L_V$ evolution for clusters with initial mass $\mcli=10^6$~\msun, including all other stellar remnants, a dissolution timescale $t_0=3$~Myr {($\tdis=17$---18.5~Gyr)}, metallicity $Z=0.02$ and a Kroupa IMF. Similar to Fig.~\ref{fig:fxsr}, solid curves denote the evolution for clusters {including the preferential loss of low-mass stars} with all neutron stars and black holes retained, long-dashed curves for clusters without these massive remnants. For clusters without {the preferential loss of low-mass stars}, short-dashed curves represent the case in which neutron stars and black holes are included and dotted curves represent the result where these remnants are removed upon their formation. The onset of {the preferential mode} $t_{\rm [pref}$ is marked by vertical lines in the linestyles of the corresponding model runs.
    }
\end{figure*}
\item[(4)]
Binaries are only partially incorporated in our models. Dynamically, they are included since our models are fitted to the collisional $N$-body simulations by \citet{baumgardt03}. However, these do not include a primordial binary population. Photometrically, binaries are not accounted for, because our model clusters are populated using single-star isochrones. As a result, we have no mechanism in which white dwarfs can evolve towards neutron stars due to mass transfer. If we had, it would increase the total remnant mass by a very small amount proportional to the fraction of white dwarfs undergoing mass transfer in a binary system.
\item[(5)]
The preferential loss of low-mass stars is included by monotonously increasing the minimum stellar mass $\mmin$ of the bound stars in the cluster. As is discussed in Sect.~\ref{sec:fxseg}, this is an approximation to the true evolution of the stellar mass function. The $N$-body simulations by \citet{baumgardt03} show that the slope of the IMF below a certain pivot-point mass $m_{\rm s}^{\rm piv}$ increases\footnote{Because the slope is negative this means that it becomes {\it less negative} and eventually more and more positive.}, thus exhibiting the preferential loss of low-mass stars. The maximum stellar mass $\mmax$ is reduced by stellar evolution. Together, these effects narrow the mass function. {In Fig.~\ref{fig:mlcomp}, our model mass and bolometric mass-to-light ratio evolution are compared to \citet{baumgardt03}. The mass evolution shows good agreement, with a small offset at intermediate age that can be attributed to the different stellar evolution prescriptions of both models (Padova 1999 for the present paper and \citet{hurley00} for \citet{baumgardt03}). Consequently, both models also differ in metallicity, as the value used by \citet{baumgardt03} is unavailable in the Padova isochrones. The difference in mass evolution is fully accounted for by these differences.}
\begin{figure*}[t]
\resizebox{\hsize}{!}{\includegraphics{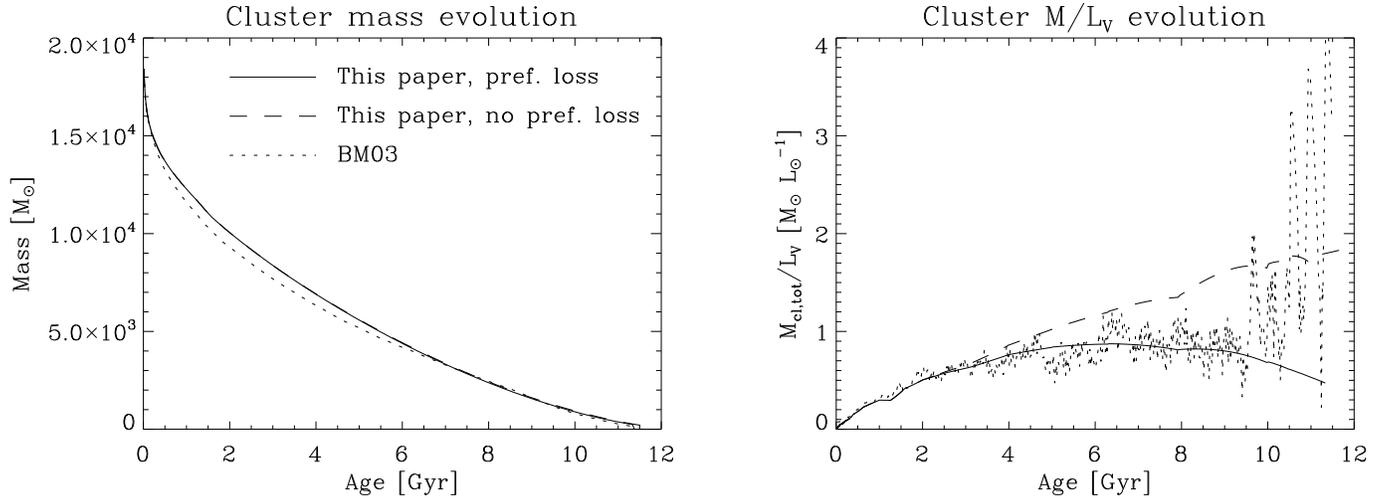}}
\caption[]{\label{fig:mlcomp}
      Comparison of the mass (left) and bolometric mass-to-light ratio evolution (right) from two of our models {with and without the preferential loss of low-mass stars} ($\mcli=18407.6~\msun$, with remnants, $t_0=22.5$~Myr ($\tdis\sim 11.5$~Gyr), $Z=0.004$ and a Kroupa IMF between 0.1~\msun and 15~\msun) to the results from \citet{baumgardt03} (same initial mass, $W_0=5$, $R_{\rm gc}=8.5$~kpc, circular orbit, $Z=0.001$ and a Kroupa IMF between 0.1~\msun and 15~\msun).
    }
\end{figure*}

{The bolometric mass-to-light ratio evolution is very similar {to our model including the preferential loss of low-mass stars} for the largest part of cluster lifetime.} A strong difference between our approach and the $N$-body simulations only arises in cluster photometry and remnant loss close to total cluster disruption, when $\mmin$ and $\mmax$ nearly meet. In that case, model clusters only retain their giant branch, because increasing $\mmin$ first removes the main sequence before giants are lost. This leads to luminosities that are overestimated close to total disruption. Moreover, this could cause real clusters to retain more remnants close to total disruption than is computed in our models. Consequently, the true mass-to-light ratio of clusters near total disruption would be larger than shown in this study. This is important in the last $\sim 15$\% of total cluster lifetime (see Fig.~\ref{fig:mlcomp}), during which the mass-to-light ratio evolution can be expected to have a positive slope due to the retain of stellar remnants rather than the computed negative one. Therefore, the slight upturn near total disruption induced by the inclusion of remnants that was shown in Fig.~\ref{fig:fxsr}(d) can be expected to be much stronger in reality. {Even a slight upturn does not appear in Fig.~\ref{fig:mlcomp} because the IMF used by \citet{baumgardt03} is truncated at 15~\msun, implying that there is no production of black holes, which are needed for the upturn to occur in models where the preferential loss of low-mass bodies is represented by an increasing lower mass limit.}

{However, it is important to note that $t=0.85\tdis$ in Fig.~\ref{fig:mlcomp} is also the point from where on the $N$-body simulations are strongly affected by statistical noise. As the number of stars in the cluster decreases and it starts to be dominated by both high- and low-luminosity objects (giants and massive remnants) the imprint of statistics on especially the mass-to-light ratio is enhanced. Consequently, it is very difficult to accurately describe the mass-to-light ratio evolution in the very last part of cluster lifetime, as such predictions are scatter-dominated and are thus  not very likely to apply to any specific real cluster.}

{The fractional range near total cluster disruption for which our models do not follow the mass-to-light ratio rise from \citet{baumgardt03} increases with the total disruption time of a cluster, with 15\% being the typical value for a $\tdis$ of about a Hubble time. Shorter lived clusters are covered for by our models for a larger part of their lifetime, while for very long-lived clusters still at least 70\% of their lifetime is covered. The vast majority of globular clusters is presently in the range where they can be treated with our models.}
\item[(6)]
We adopt a step-function form for $f_{\rm pref}(t)$ to describe the fraction of the cluster mass loss occurring in the form of low-mass stars if the cluster has (partially) {reached energy equipartition}. In the increasing minimum stellar mass approximation, a step-function requires the form of Eq.~\ref{eq:fseg!} to most accurately reproduce the $N$-body simulations by \citet{baumgardt03}. However, cluster magnitude evolution will be better reproduced if a smooth function of $t$ is formulated. Nonetheless, this is not likely to lead to an exact or better representation of \citet{baumgardt03} due to the fundamentally different approach of including the preferential loss of low-mass stars. Instead, in a future study we will incorporate an improved description of the changing mass function due to the loss of low-mass stars to account for the effect of {the preferential loss of low-mass stars} (see Sect.~\ref{sec:fut}).
\item[(7)]
Stellar remnants with masses smaller than the minimum stellar mass in the cluster, i.e., $\mssr<\mmin$ are immediately available for dissolution. {If the escape rate of bodies is not constant with cluster radius}, this ignores the outward transport of remnants from their birth location (for a mass-segregated cluster this is considered to be in the cluster centre) to the cluster outskirts on the half-mass relaxation timescale $t_{\rm rh}$. Because generally $t_{\rm rh}<\tdis-t$ except close to total cluster disruption (see Appendix~\ref{sec:app}), this is a reasonable approach. In the exceptional case where $t_{\rm rh}>\tdis-t$, remnant loss is halted since the cluster is disrupted on a shorter timescale than $t_{\rm rh}$.
\item[(8)]
{The speed of cluster dissolution is assumed to be independent of cluster radius. \citet{gieles08} have shown that this is a reasonable assumption for most tidally dissolving clusters, especially for large-$N$ systems like globular clusters.}
\end{itemize}

\subsection{Applicability and future studies} \label{sec:fut}
Because this study is based on collisional $N$-body simulations that confirm the existence of {the preferential loss of low-mass stars} and the retain of stellar remnants, the predicted effects will be present in real clusters. It is shown that these phenomena, but also IMF and metallicity variations can have unique effects on either the mass, magnitude, colour and mass-to-light ratio evolution of clusters. Therefore, the effects of {the preferential loss of low-mass stars}, stellar remnants, IMF and metallicity can be expected to be observable and interpretable. Clusters of all ages between 10~Myr and 19~Gyr can be treated with our models. Near total cluster disruption the results are affected by our formulation of the preferential loss of low-mass stars, in which red giants are the very last bodies to be lost near total cluster disruption. As a result, the cluster magnitude is overestimated and the mass-to-light ratio is underestimated during the last $\sim 15$\% of cluster lifetime. Therefore, photometry-related observables have to be used with caution near total cluster disruption (see Sect.~\ref{sec:assump}).

The metallicity dependence of stellar remnant mass and an improved description of the preferential loss of low-mass stars are to be included in future studies. This will provide more accurate descriptions of non-luminous cluster mass, and cluster photometry and mass-to-light ratio close to total disruption. {A new set of evolutionary synthesis models with a time-dependent stellar mass function, based on $N$-body simulations will soon be available \citep{anders08}.} Moreover, we have made a quantitative comparison of our models to a number of globular cluster systems, and have assessed unexplained features of the mass-to-light ratio distribution \citep{kruijssen08b}. {A paper in which the individual properties of Galactic globular clusters, like their orbital parameters and metallicities, are used to explain their mass-to-light ratios is in preparation \citep{kruijssen08c}}.

\section{Conclusions} \label{sec:concl}
We have treated the influence of {the preferential loss of low-mass stars}, stellar IMF, metallicity and the inclusion of stellar remnants on cluster mass, magnitude, colour and mass-to-light ratio evolution. We presented analytical models that describe the evolution of cluster content and photometry, based on stellar evolution from the Padova 1999 isochrones and on simplified dynamical dissolution models as first presented in \citet{lamers05}. The latter, in turn, is based on the $N$-body simulations by \citet{baumgardt03}. 

{The models represent the cluster evolution part of our new cluster population synthesis code {\it SPACE}.} We considered Kroupa and Salpeter IMFs and metallicities in the range $Z=0.0004$---0.05. The obtained data are publicly available in electronic form at the CDS and also at \texttt{http://www.astro.uu.nl/\~{}kruijs}. The results from our models are as follows.
\begin{itemize}
\item[(1)]
{\it The preferential loss of low-mass stars} slightly decreases the total disruption time of a cluster by a few percent. However, the most significant changes are effected in cluster photometry. The effect of fading is decreased as clusters {including mass loss in the preferential mode} can stay more than 1.5 $V$-band magnitudes brighter than clusters losing mass in the canonical mode, because most of the dynamical mass loss occurs in the form of low-mass stars that contribute little to cluster luminosity. Initially, clusters {exhibiting the preferential loss of low-mass stars} are bluer than standard ones, but they become redder during the last $\sim 10\%$ of cluster lifetime. The cluster mass-to-light ratio is severely decreased due to {the preferential loss of low-mass stars}. The decrease typically ranges from 2---4$~\msun~{\rm L}_\odot^{-1}$ (i.e., up to 0.6~dex) near total cluster disruption {for total disruption times $\tdis>12$~Gyr}. {If the upturn of the $M/L_V$ evolution that is much more prominent in \citet{baumgardt03} than in our models (see Sect.~\ref{sec:assump}, point (5)) is accounted for, this range of $M/L_V$ decrease is at most 0.5$~\msun~{\rm L}_\odot^{-1}$ smaller.}
\item[(2)]
{Including the mass of {\it stellar remnants} obviously yields an increase} in the total cluster mass and consequently also in total disruption time with respect to cluster evolution without remnants. The extended lifespan also implies that cluster luminosity less rapidly decreases. The mass-to-light ratio is enhanced by almost 2 $~\msun~{\rm L}_\odot^{-1}$ at its maximum, close to total disruption.
\item[(3)]
We compared the evolution of clusters with {\it Salpeter and Kroupa IMFs}, which can be considered to {favour high stellar masses (Kroupa, or a `top-oriented' IMF) or low masses (Salpeter, or a `bottom-oriented' IMF) alternatives due to the bend in the Kroupa IMF and a slight slope difference}. As can be expected, clusters with a bottom-{oriented} IMF retain more mass due to stellar evolution, which eventually causes these clusters to become brighter than clusters with a top-{oriented} IMF. However, they start out being slightly fainter since a top-{oriented} IMF favours massive stars and is thus brighter than a bottom-{oriented} one. Similarly, clusters with a top-{oriented} IMF are bluer and have smaller mass-to-light ratios than clusters with a bottom-{oriented} IMF. {For the Kroupa and Salpeter IMFs}, the latter change can amount up to several $~\msun~{\rm L}_\odot^{-1}$.
\item[(4)]
{\it Metallicity} variations hardly influence the total mass evolution of clusters. In accordance with stellar studies \citep{hurley04} low-metallicity clusters are brighter and also much bluer than high-metallicity ones. Consequently, the mass-to-light ratio is a strongly increasing function of metallicity.
\item[(5)]
When applying our results to {\it Galactic globular clusters}, it is evident that {the preferential loss of low-mass stars} is required to explain their low observed {\it mass-to-light ratios}, especially if stellar remnants are accounted for. Low metallicity is insufficient to serve as an explanation. Another important implication of our study is that the mass-to-light ratio can {\it not} be assumed to be constant over varying luminosity, as it is strongly affected by the dynamical history of clusters.
\item[(6)]
The fact that clusters of high masses may not have reached energy equipartition yet suggests that the effects of {the preferential loss of low-mass stars} disappear with increasing cluster mass. Because clusters {exhibiting the preferential loss of low-mass stars} have much lower mass-to-light ratios than clusters that lose their mass in the canonical mode, clusters with high masses would then have much higher mass-to-light ratios than ones with lower masses. This effect may have been found by \citet{rejkuba07}. The above interpretation and its application to the observations of \citet{rejkuba07} is treated more extensively in \citet{kruijssen08b}.
\item[(7)]
The typical {\it colour range of globular clusters} is covered by our models. When considering the colour-metallicity relation as reported by \citet{smith07}, from an order-of-magnitude comparison we suggest that the observed colour scatter at fixed metallicity could be the effect of {the preferential loss of low-mass stars}.
\item[(8)]
{Only when adopting a Salpeter IMF down to $\mmini=0.1$~\msun, the mass-to-light ratios of UCDs are reproduced by our models. While UCDs could represent a natural continuation of the trend of increasing mass-to-light ratio with (globular) cluster mass \citep{rejkuba07}, this is not expected to be of a dynamical nature, since more massive UCDs are not expected to have reached energy equipartition within a Hubble time.}
\end{itemize}

The retain of remnants and the existence of {the preferential loss of low-mass stars} are found in $N$-body simulations of clusters and in observations, while metallicity and IMF variations are observed among real clusters. Therefore the effects described in this paper should be considered when observing clusters, and observed cluster properties have to be interpreted very carefully\footnote{Predictions for specific models can be made by the first author upon request.}.

\section*{Acknowledgments}
We thank the International Space Science Institute (ISSI) in Bern, Switzerland, for hosting a star cluster workshop where part of this project was carried out. We are grateful to Holger Baumgardt for {comments on this paper and for} kindly providing the data from \citet{baumgardt03}. We also thank {Mark Gieles, Marina Rejkuba, Steffen Mieske, Dean McLaughlin,} Reinier Zeldenrust and Peter Anders for valuable discussions, comments and suggestions. {The anonymous referee is acknowledged for constructive comments that strongly improved the manuscript.}

\bibliographystyle{aa}
\bibliography{mybib}

\appendix
\section{Outward motion of stellar remnants} \label{sec:app}
In this study it is assumed that when stellar remnants are the lowest mass bodies in a cluster {that preferentially loses low-mass stars} they are immediately available for dissolution. In mass-segregated clusters, remnants are created in the cluster centre, where the most massive stars reside. Because bodies are only lost from the cluster if they cross the tidal radius, this leads to a delay compared to remnants that would be produced at all radii {unless the escape rate from the cluster is independent of radius as proposed by \citet{king66}. In this Appendix we show that the delay can be neglected even if the escape rate varies throughout the cluster.}

If the lowest mass bodies in the cluster are stellar remnants, these will move outwards on a half-mass relaxation timescale $t_{\rm rh}$ \citep[e.g.,][Ch.~14]{spitzer87,heggiehut}. Our description of remnant loss ignores any delay caused by the motion of remnants from the cluster centre to its outskirts. Therefore it implicitly assumes $t_{\rm rh}\ll \tdis-t$, where the latter term represents the remaining lifetime of the cluster. To compare the two terms, we define $\chi=(\tdis-t)/t_{\rm rh}$. If $\chi>1$, remnants are able to reach the tidal radius before total cluster disruption; for $\chi<1$, the cluster is completely disrupted before such equillibrium can be reached. 

\begin{figure}[t]
\resizebox{\hsize}{!}{\includegraphics{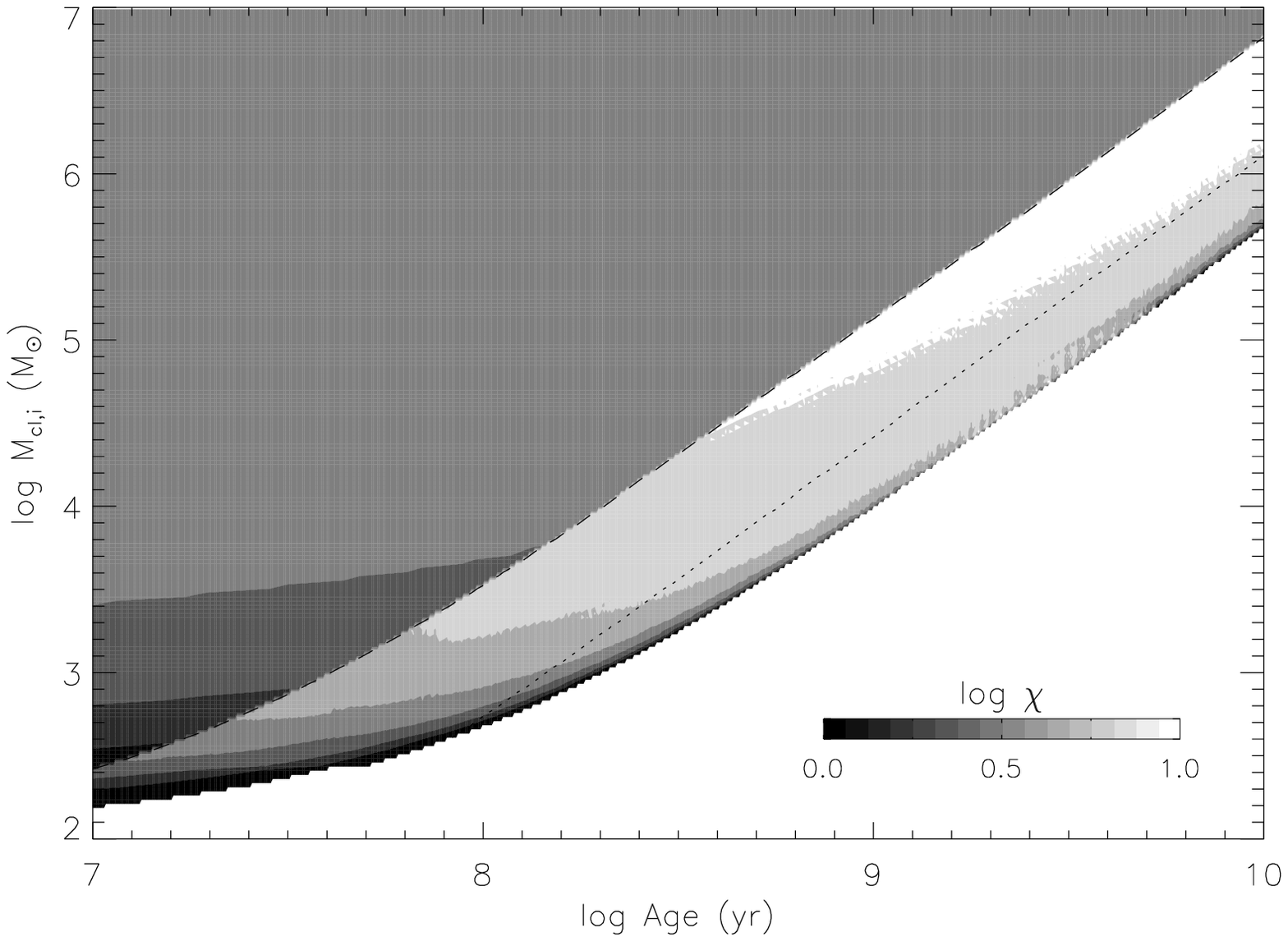}}
\caption[]{\label{fig:trh}
      Logarithm of the ratio of remaining cluster lifetime and the half-mass relaxation timescale, $(\tdis-t)/t_{\rm rh}\equiv\chi$, as a function of cluster age $t$ and initial cluster mass $\mci$. The dashed line represents the {onset of the preferential mode} $t_{\rm pref}$ for a cluster of corresponding y-axis initial mass, and the dotted line denotes the time from which on remnants are lost from such a cluster ($t_{\rm sr}$, see Sec.~\ref{sec:dissseg} for details). For the displayed model run we used a dissolution timescale of $t_0=3$~Myr, metallicity $Z=0.02$, a Kroupa IMF and complete {energy equipartition} for $t\geq 0.2\tdis$.
    }
\end{figure}

In Fig.~\ref{fig:trh}, $\log{\chi}$ is shown for initial cluster masses between $10^2$ and $10^7$~\msun, with a dissolution timescale of $t_0=3$~Myr, metallicity $Z=0.02$, a Kroupa IMF and complete {energy equipartition} ($f_{\rm pref}(t)=1$) after $t=0.2\tdis$. When a cluster reaches {energy equipartition}, there is a rapid drop in $t_{\rm rh}$ because the mean mass in the cluster centre increases. This is reflected in a sudden increase of $\log{\chi}$ with age for any specific initial mass. This can be observed in Fig.~\ref{fig:trh} at the dashed curve, which represents the onset of {the preferential mode} $t_{\rm pref}$ (see Sect.~\ref{sec:fxseg}) for the entire initial mass range. From Fig.~\ref{fig:trh} and model runs for other choices of dissolution timescale, metallicity and IMF, we can conclude that after $t_{\rm pref}$, $\chi \simgreat 3$ for all parameter sets relevant to globular clusters, and still $\chi \simgreat 1.4$ for clusters with initial masses $\mci<10^3~\msun$ and extremely rapid cluster dissolution ($t_0=0.3$~Myr). For massive clusters, we typically have $\chi\sim 10$ for $t>t_{\rm pref}$, implying that the immediate availability of remnants is a legitimate approximation.

Only during the very last few Myrs before total cluster disruption a cluster can have $\chi<1$ because the numerator ({\it remaining} cluster lifetime) approaches zero more rapidly than the denominator ($t_{\rm rh}$). Remnants that are produced during that brief phase cannot be lost in the {preferential} mode. If just before total disruption $\chi<1$, the lower integration limit $\mmax$ in Eq.~\ref{eq:mclsrav}, that determines the total remnant mass available for dissolution, should be replaced by $m_{\rm max}(t_{\rm\chi})$, with $t_{\rm\chi}$ the time at which $\chi$ decreases below unity\footnote{To determine $\chi$, the total disruption time $\tdis$ has to be known before having integrated cluster properties over (and thus having obtained-) the entire cluster lifetime. This is alleviated by estimating $\tdis$ using straightforward integration of Eq.~\ref{eq:dmdt} at the very start of the computation.}.

As we have shown, for other parts of cluster evolution the assumption of the immediate loss of remnants with masses $\mssrt\leq \mssr\leq\mmin$ is reasonable. Please note that for clusters {including the preferential loss of low-mass stars} with $t\geq t_{\rm pref}$, typically $t_{\rm rh}$ is not only very small compared to the remaining lifetime of the cluster, but also ($\simless 10\%$) compared to cluster age.

\end{document}